%% file: base.tex
\documentclass[12pt]{iopart}

\usepackage{graphicx}
\usepackage{verbatim}
\usepackage{color}
\usepackage{subfigure}
\usepackage{wasysym}
\usepackage{cite}

\begin{document}

\title{Mott Insulators in a Fully-Frustrated Bose Hubbard Model on the Honeycomb Lattice}

\author{Stephen Inglis and Roger G. Melko}
\address{Department of Physics and Astronomy, University of Waterloo, Ontario, N2L 3G1, Canada}

\newcommand{\needcite}{{\color{red}[NEED CITATION]}}
\newcommand{\err}[1]{{\color{red}[#1]}}

\input{abstract.tex}
\input{introduction.tex}

\input{results.tex}

\input{conclusions.tex}

\bibliography{rand}{}
\bibliographystyle{unsrt}

\end{document}

%% file: abstract.tex
\begin{abstract}
We examine the effects of quantum fluctuations on a classical spin liquid state in the fully-frustrated honeycomb lattice Bose Hubbard model using quantum Monte Carlo simulations.  Frustration is induced explicitly in the model by modulating the sign of the interaction spatially around each lattice hexagon.  A superfluid to Mott insulating quantum phase transition can be induced by varying the relative strength of the classical interaction and quantum hopping.  In the cases where the interaction has a regular spatial modulation, hopping promotes a phase transition to a symmetry-broken valence-bond solid state.  When the interaction is forced to have no regular pattern, the Mott insulating phase is found to be featureless and gapped, making it an interesting candidate state for a quantum spin liquid arising in a Hamiltonian with only nearest-neighbor interactions.
\end{abstract}

%% file: introduction.tex
\section{Introduction}

A concentrated effort is currently underway to build exotic states of quantum matter in ultracold atomic systems~\cite{greiner2008condensed, bloch2008, recati2003, greiner2002quantum, jaksch1998}.  These atomic systems are destined to become ``quantum simulators'' of condensed matter Hamiltonians, allowing us to extend our reach into new models, phases, and phase transitions inaccessible to traditional condensed matter theory or experiment~\cite{greiner2002quantum, burkov2006}.  Although efforts to construct complex Bose or Fermi Hubbard Hamiltonians is still a fledgling effort, experimental progress is rapid, motivating theorists to continue to study increasingly realistic models that may someday be built and used in the search for exotic quantum phenomena.

Concomitant with this revolution, there has been a resurging interest in frustrated systems recently, due to the possibility that 
when frustration suppresses the ordering tendencies of a system, more subtle exotic phenomena can be observed.
Of particular success would be the stabilization of a quantum spin liquid~\cite{RVB1,RVB2,balents2010spin, wen_classify} -- a featureless Mott insulating (MI) phase with emergent gauge symmetry that harbors fractional excitations -- in a cold atom system.  Spin liquids are projected to occur in several two-dimensional Bose Hubbard models~\cite{isakov_spinliquid, leon_kagome}, however the typical feature shared by these Hamiltonians is that the range of the kinetic hopping (tunneling) and interactions are multi-particle, long-range, or otherwise prohibitively complicated for realization in real experimental setups.

In this paper, we construct a featureless Mott insulating phase from a Bose-Hubbard Hamiltonian with only nearest-neighbor interactions.  Motivated by the resonating valence-bond (RVB) \cite{RVB1,RVB2} phase contained in the triangular lattice quantum dimer model \cite{moessner_qdm}, our model is equivalent to a spin-1/2 XXZ model constructed by adding quantum XY (hopping) perturbations on top of a fully-frustrated Ising model on the honeycomb lattice.  We find that the hopping term lifts the degeneracy of the classical Ising model ground state, promoting various Valence Bond Solid (VBS) phases (with coexisting charge-density wave order), depending on the spatial distribution of the lattice frustration.  Only when the spatial position of the frustration is made random, essentially breaking all lattice symmetries, does an incompressible (gapped) Mott insulator, a candidate quantum spin liquid phase, appear in the low-temperature phase diagram of the model.

\section{Fully-Frustrated Honeycomb Lattice Bose Hubbard Model}

In a recent breakthrough, Becker {\it et.~al.}~\cite{Becker} have constructed a versatile experimental setup for loading ultracold atoms into a triangular optical lattice, demonstrating for the first time a superfluid-MI transition in this 2D system.  Their experimental realization involves creating a periodic potential constructed from three laser beams in the XY plane.  A rotation of the polarization of the three lasers in the XY plane configures the potential minima in the geometry of a honeycomb lattice.  In the typical unfrustrated case, this bipartite lattice is expected to induce antiferromagnetic ordering for atoms \cite{wessel}.  However, Becker {\it et.~al.}'s demonstration that the {\it sign} of the tunneling matrix element can be reversed may suggest the ability to induce frustrated or anisotropic tunneling of the atoms between lattice sites.  It may therefore be possible in the future to introduce frustration into otherwise geometrically unfrustrated models.  We explore this scenario by examining the possible phases of the simplest fully-frustrated honeycomb lattice Bose Hubbard model with nearest-neighbor interactions using large-scale QMC simulations.

We consider very generally the simplest model for hard-core bosons on the honeycomb lattice:
\begin{equation}
H =  \sum_{ij} \Bigl( J_{ij}(n_i - 1/2)(n_j - 1/2) \Bigr) - t/2 \sum_{ij} \Bigl( b^\dagger_i b_j + b_i b^\dagger_j \Bigr), \label{BoseHam}
\end{equation}
employing only nearest-neighbor hopping ($b^\dagger_i$ and $b_i$ are the boson creation and annihilation operators), and density-density interactions ($n_i=0$ or 1).
Here, the density-density interaction is written such as to allow for the exact mapping to the quantum spin-1/2 XXZ Hamiltonian,
\begin{eqnarray}
H = \sum_{ij} \Bigl( J_{ij} S^z_i S^z_j \Bigr) - t \sum_{ij} \Bigl( S^x_i S^x_j + S^y_i S^y_j \Bigr), 
\end{eqnarray}
via $ S_i^+ \rightarrow b_i^\dagger$, $S_i^- \rightarrow b_i$, and $n_i \rightarrow S_i^z+1/2$.
Considering first the limit when $t=0$, one has the classical Ising model, where frustration maybe be induced by ensuring that the {\it sign} of $J_{ij}$ is modulated in such a way that nearest-neighbor interactions can not be fully satisfied around a hexagonal plaquette. 
In the extreme limit where each hexagon on the lattice is frustrated (i.e.~the model is ``fully-frustrated'') the low-temperature phase is an extensively degenerate manifold of equal-energy states -- a type of classical spin liquid \cite{Honey_1}. 
Each configuration in the degenerate manifold of states can be mapped to a close-packed hard-core dimer configuration on the dual triangular lattice~\cite{Tri_dimer}.  It is well-known that the quantum version of the triangular lattice dimer model contains an RVB liquid in its phase diagram \cite{moessner_qdm}. Although an exact mapping from the quantum dimer model to a quantum spin model is not available, one may consider the $t=0$ classical fully-frustrated Ising model groundstate as a starting point, and ask 
whether quantum perturbations in the $t \neq 0$ model contains the spin counterpart to the RVB phase \cite{RVB1,RVB2}; i.e.~a Z2 quantum spin liquid phase \cite{moessner_qdm}.

Our Hamiltonian \ref{BoseHam} has a distinct advantage over some other frustrated spin models since it may be simulated without the sign problem using Stochastic Series Expansion (SSE) quantum Monte Carlo~\cite{SSE1,SSE2,sandvik_sse2,sandvik_sse} at finite temperature.
Imposing the fully-frustrated constraint but keeping the magnitude of the interaction isotropic on the lattice, one can re-write the model
\begin{equation}
H =  J\sum_{ij} (-1)^{A_{ij}}  (n_i - 1/2)(n_j - 1/2) - t/2 \sum_{ij} \Bigl( b^\dagger_i b_j + b_i b^\dagger_j \Bigr),
\end{equation}
where $A_{ij}$ takes the value of zero or one on each bond, subject to the constraint that the sum over each hexagon, $\sum_{\hexagon}A_{ij}$, is odd.  When $t \neq 0$ the model is {\it not} fully gauge-invariant, and modulating the sign of the hopping $t$ would introduce a sign problem; we nonetheless refer to the choice of $A_{ij}$ subject to the constraint a {\it gauge} choice.

\begin{figure}[h]
\begin{center}
\includegraphics[width=0.75\textwidth]{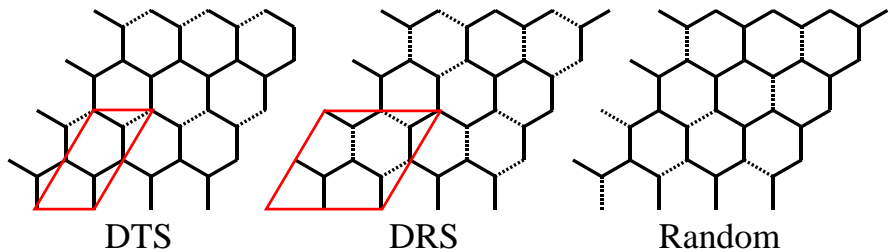}
\caption{The three gauges of study in this paper, the ``Discrete Translationally Symmetric'' (DTS) gauge, the ``Discrete Rotationally Symmetric'' (DRS) gauge and the ``random'' gauge.
Sold bonds have $A_{ij}=0$, and dashed bonds have $A_{ij}=1$.
Unit cells are marked in red, and there is no unit cell for the random gauge.
\label{fig:gauges}
}
\end{center}
\end{figure}
In the classical limit, the choice of the gauge $A_{ij}$ does not affect the nature of the groundstate manifold, due to the one-to-one mapping to the classical hard-core dimer tiling, which 
retains a residual entropy of $S = 0.214$ per site~\cite{Tri_dimer}.
In this paper, we examine three different gauge choices (each with $J$ positive and negative) to study the effect of the quantum hopping ($t$-term) on the classically degenerate manifold (see Figure~\ref{fig:gauges}).
The first choice of gauge we refer to as the ``Discrete Translationally Symmetric'' (DTS) gauge where the pattern of bonds where $A_{ij} = 1$ reduces the symmetry of the lattice such that only a discrete translational symmetry remains.
We also study a gauge we call the ``Discrete Rotationally Symmetric'' (DRS) gauge, where the $A_{ij}$ variables leave the system with a nontrivial rotational symmetry and reduced translational symmetry.
The third gauge is one in which we use a classical Monte Carlo to generate an unbiased random pattern of $A_{ij}$ that satisfies the fully-frustrated requirement $\sum_{\hexagon} A_{ij}$ =1, and this is called the ``random'' gauge.
Figure~\ref{fig:gauges} illustrates each gauge with their unit cell outlined (where a unit cell exists).

%% file: results.tex
\begin{figure}
\begin{center}
\includegraphics[width=0.75\textwidth]{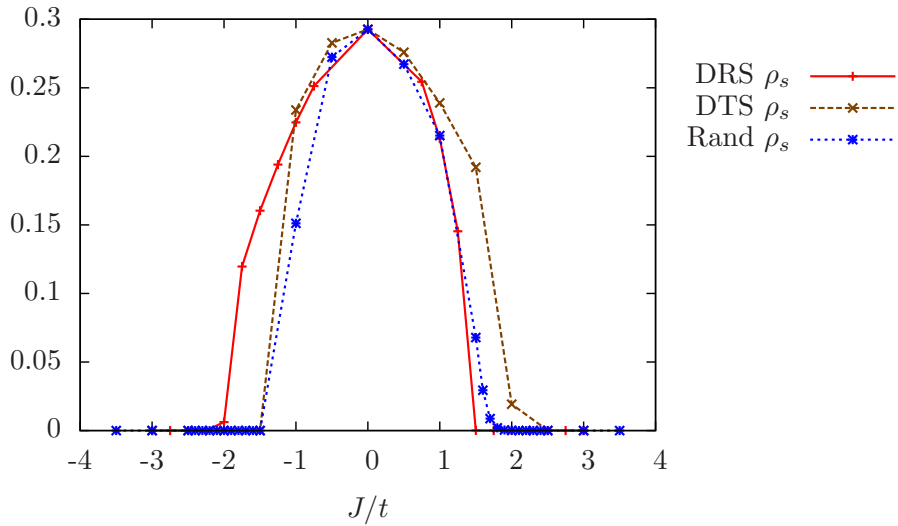}
\caption{The superfluid density (or spin stiffness), $\rho_s$, for each of the different gauge choices at $T = 0.1 t$.
\label{fig:rho}
}
\end{center}
\end{figure}

We now examine the results of QMC simulations on the models defined in the previous section.
We fix $t=1$ in all results to follow. 
Note first that when $J/t=0$ the sign of $J$ on the bonds is irrelevant, and the model becomes
an XY-superfluid for all gauge choices, 
characterized by a non-zero superfluid density (or spin stiffness) of $\rho_s = 0.292(1)$.
This is the limit where SSE QMC with directed loops works most efficiently, hence we approach the transition
to Mott insulating behavior by increasing $J/t$ from zero in the simulations.
We begin by examining the superfluid density (measured via the winding number~\cite{sse_spin_stiffness}) at a fixed low temperature, $T = 0.1 t$, as a function of increasing $J/t$.  
Results are illustrated in Figure~\ref{fig:rho} for the three gauges.
We observe that quantum phase transitions between the superfluid and various insulating states
are promoted by an interaction $|{J_c}| \approx 2$.  The specific value of the critical $J_c$ 
is similar, but not strictly identical for the different gauges, or for the opposite signs of $J$ in a fixed
gauge.

Using the three different gauge choices, we now have access to six different insulating states (including $\pm J$).  
In order to determine the ordering nature of the ground state, if any, we first examine $q$-space structure factors
at a fixed $J$ and temperature $T$.  $|J|=3$ was chosen to be well in the insulating phase, but of sufficiently small
interaction to ensure algorithmic ergodicity.  The $T$ dependence of most observables was examined to find the
approximate low-T convergence: $T=0.1 t$ was chosen in most of the below data.  
Also, although the SSE QMC is in general very efficient at finding the ground state of a system, it is also capable of becoming stuck in a local minima in configuration space, similar to classical Monte Carlo algorithms.
An example of such a local minima would be a screw dislocation if the ground state pattern were broken into a set of strips.
The energy cost of such a defect would be finite and although translating the defect should be possible, fixing it requires passing through intermediate states of even higher energy.
In order to overcome this problem we use the technique of thermal annealing.
We start the simulation at a high temperature, run a fixed number of Monte Carlo steps (enough to equilibrate) and then lower the temperature until the desired temperature is reached.
By using this method we generate low temperature configurations that tend to be free of defects.

\begin{figure}
\begin{center}
\subfigure[DTS, $J=-3$]{
\includegraphics[width=0.4\textwidth]{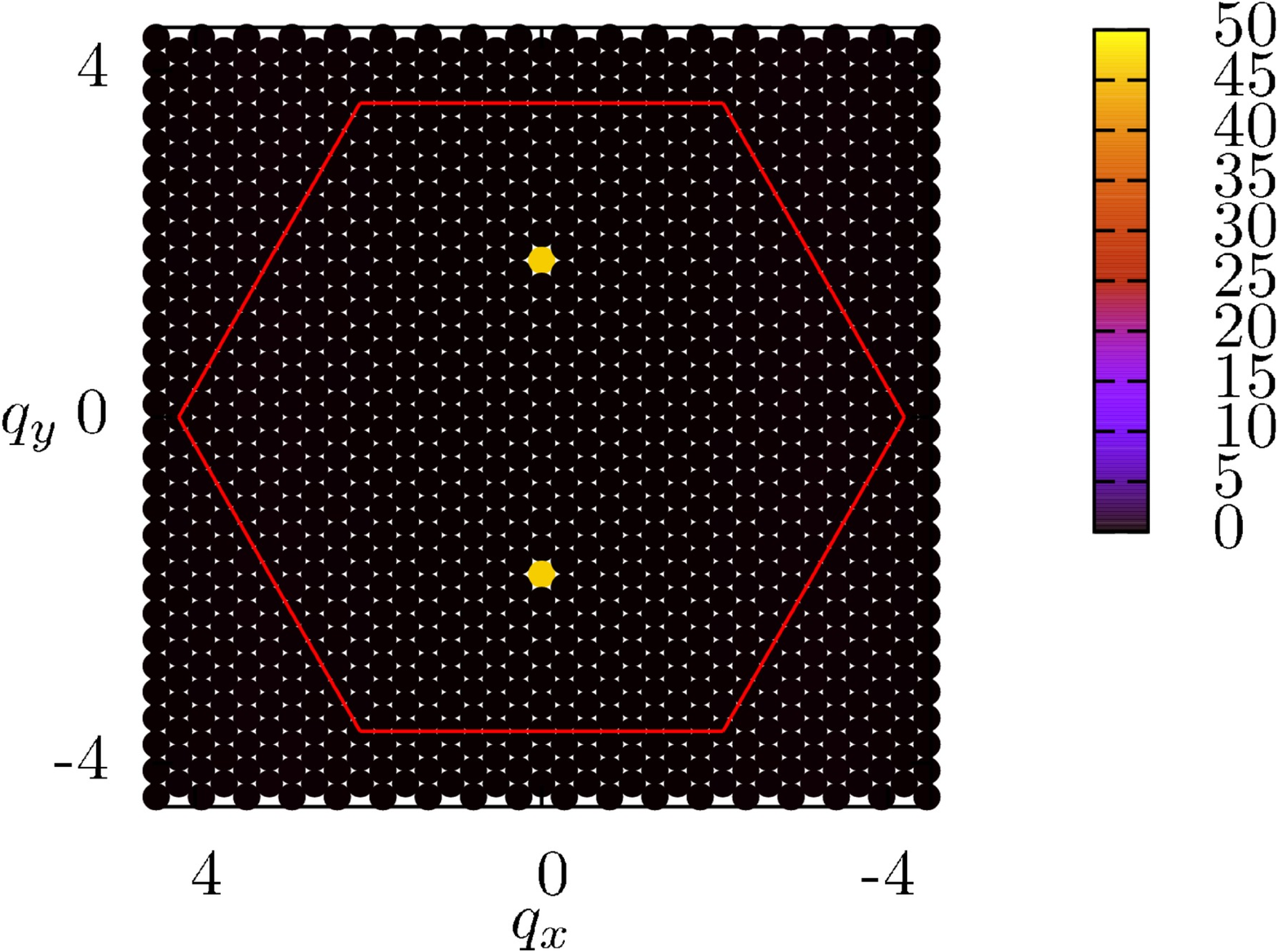}
\label{fig:spinstructure:reg_fm}
}
\subfigure[DTS, $J=3$]{
\includegraphics[width=0.4\textwidth]{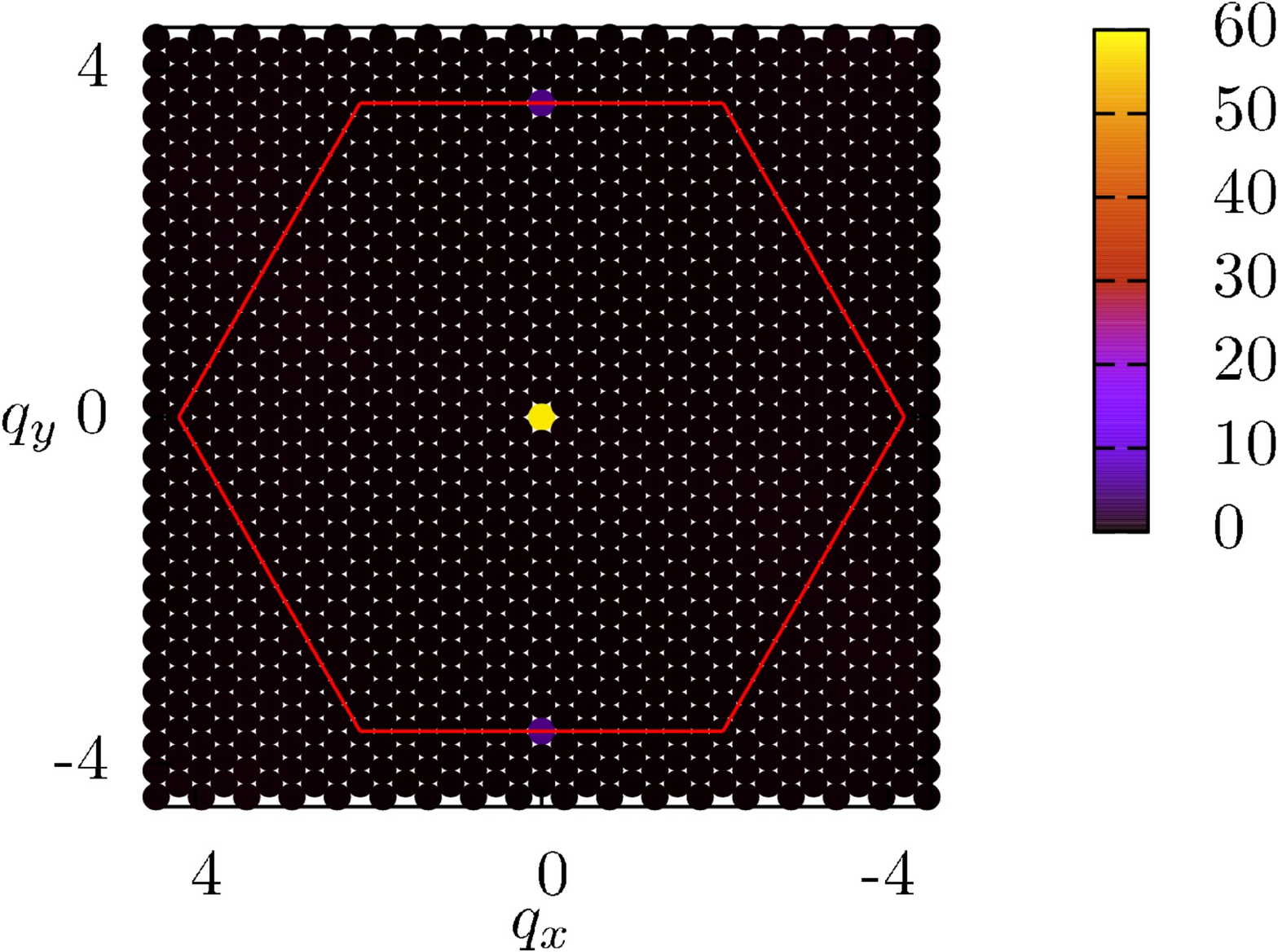}
\label{fig:spinstructure:reg_afm}
}
\subfigure[DRS, $J=-3$]{
\includegraphics[width=0.4\textwidth]{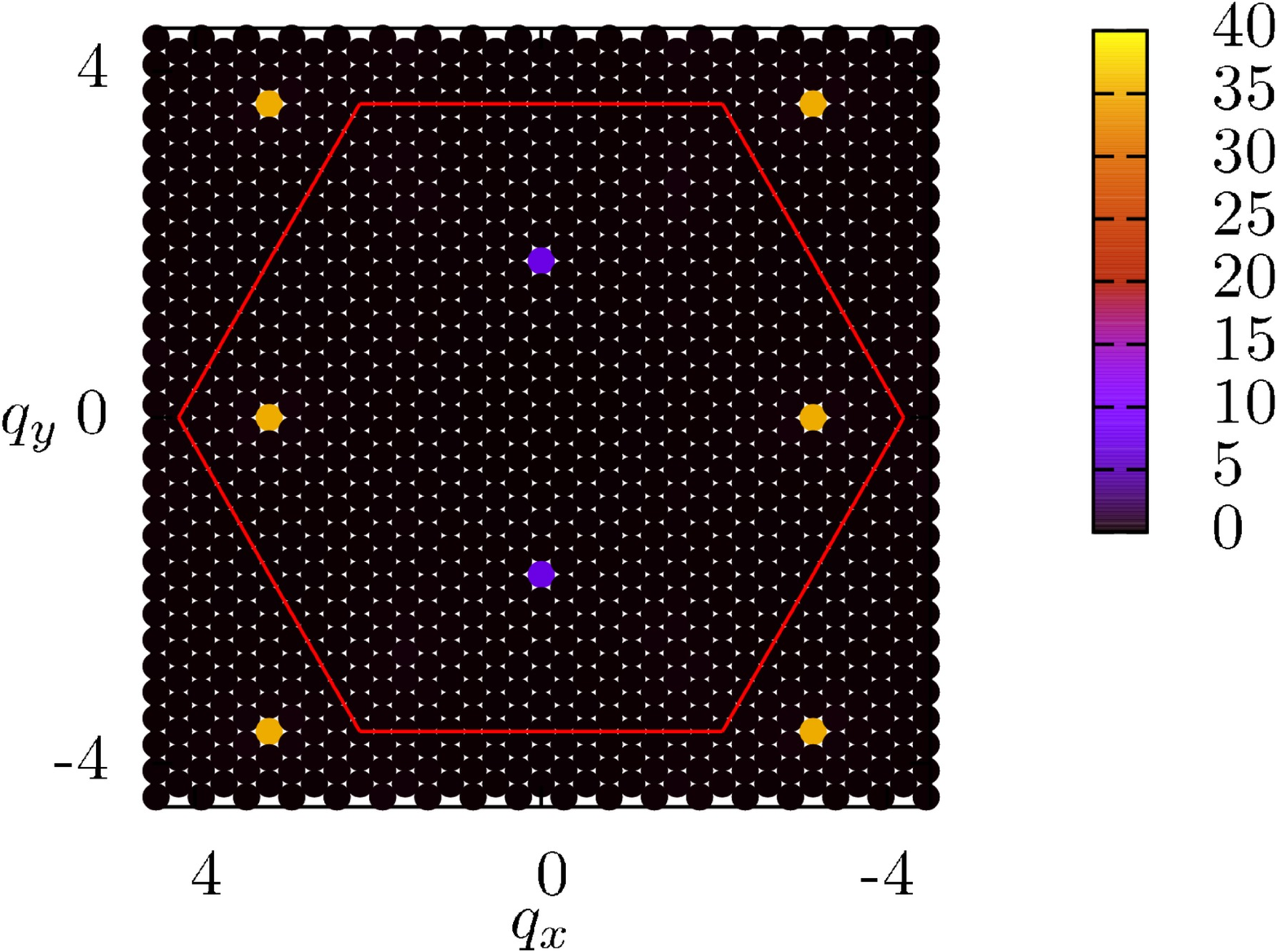}
\label{fig:spinstructure:symm_fm}
}
\subfigure[DRS, $J=3$]{
\includegraphics[width=0.4\textwidth]{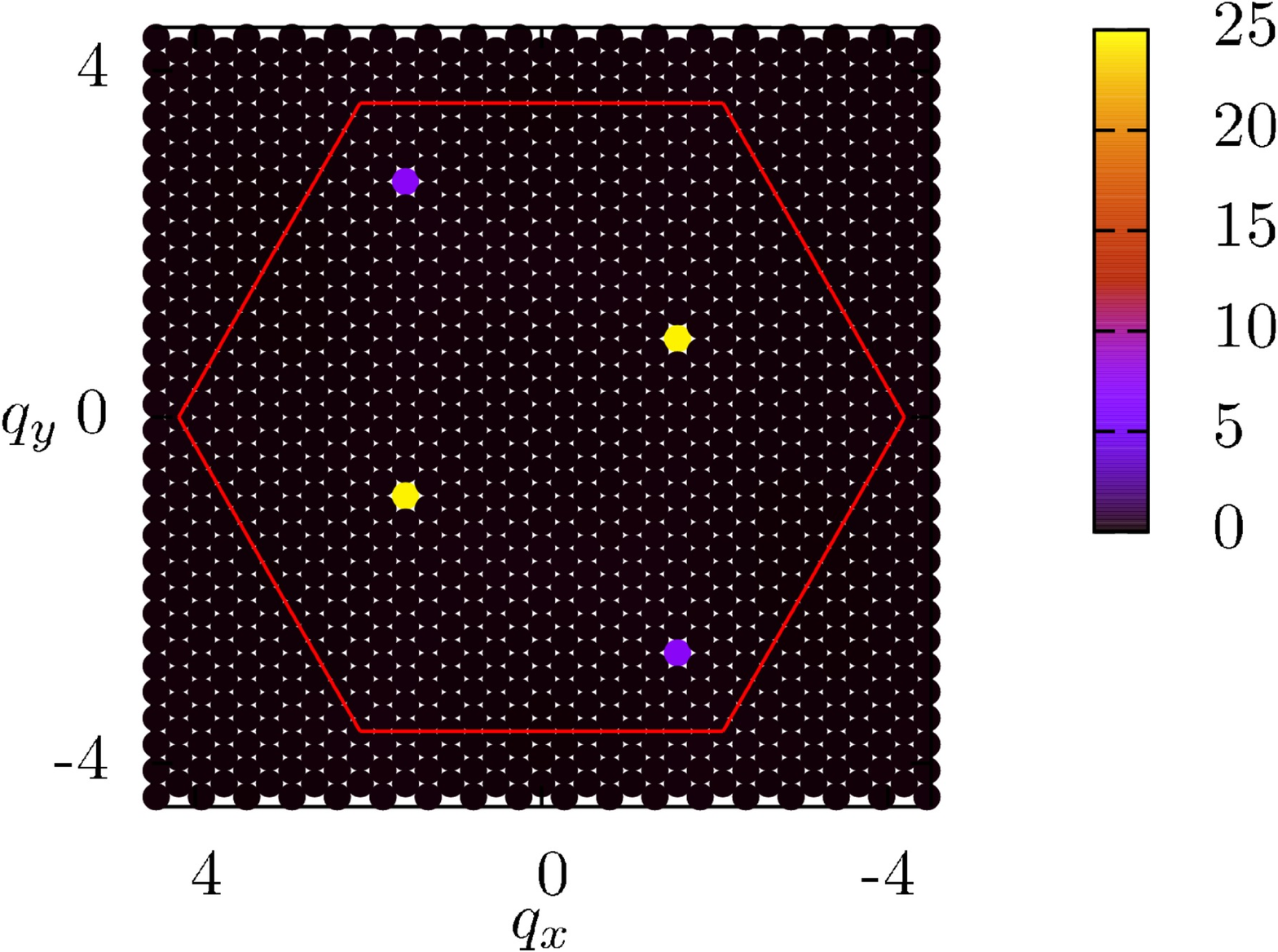}
\label{fig:spinstructure:symm_afm}
}
\subfigure[Random, $J=-3$]{
\includegraphics[width=0.4\textwidth]{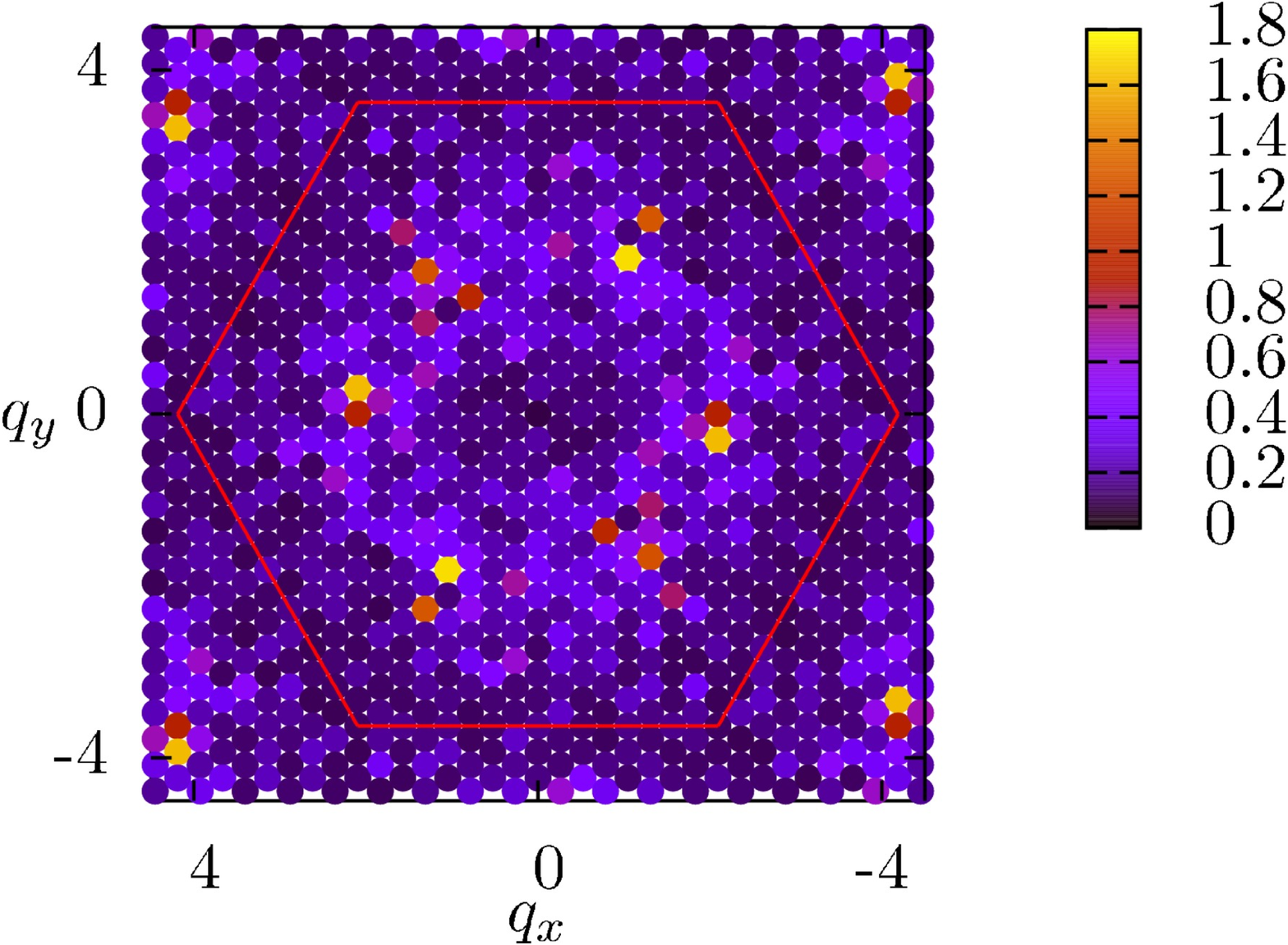}
\label{fig:spinstructure:rand_fm}
}
\subfigure[Random, $J=3$]{
\includegraphics[width=0.4\textwidth]{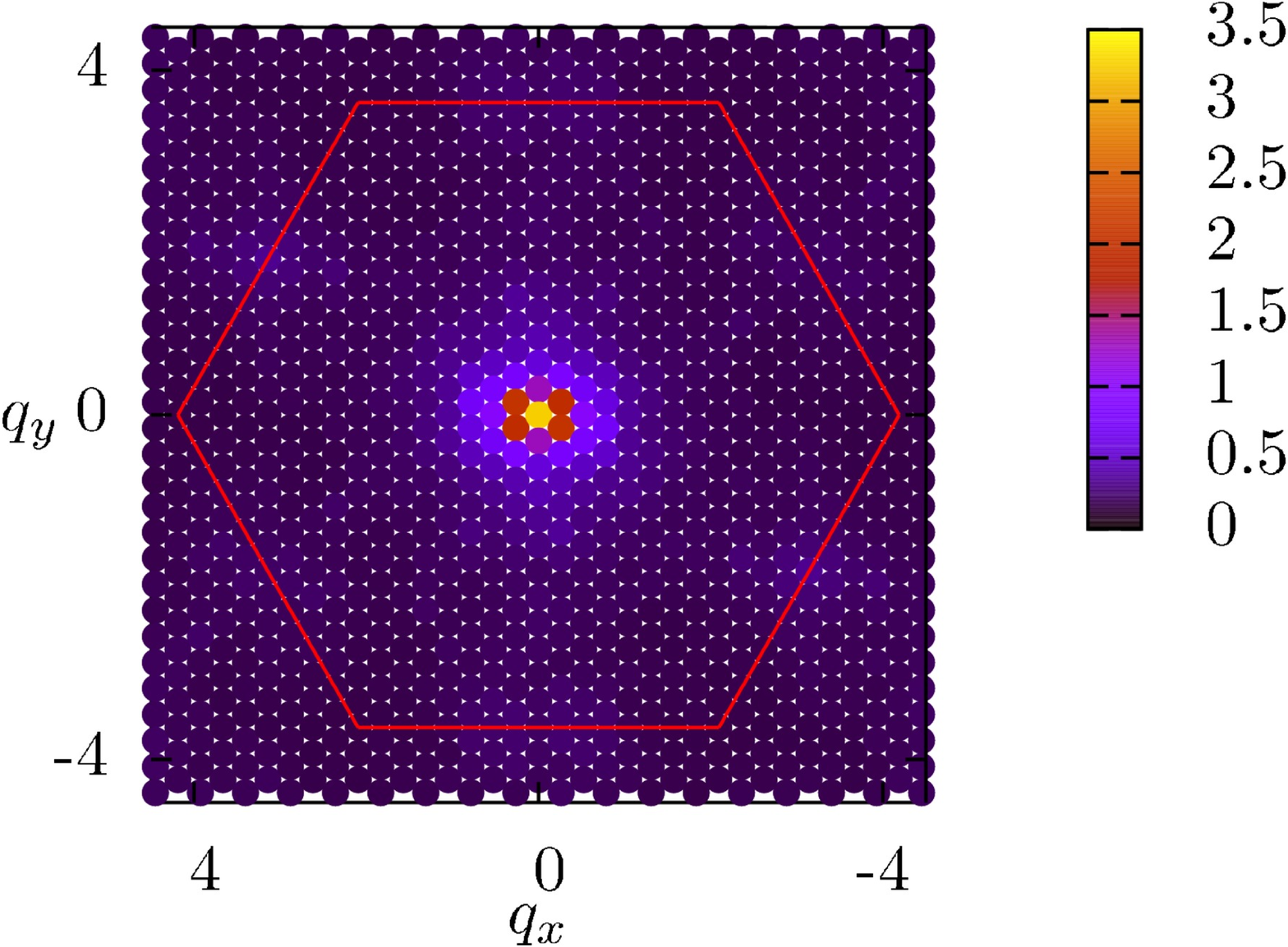}
\label{fig:spinstructure:rand_afm}
}
\caption{The structure factor of the density-density correlation function for the six cases studied in this paper.
The peaks in~\subref{fig:spinstructure:reg_fm},\subref{fig:spinstructure:reg_afm},\subref{fig:spinstructure:symm_fm},\subref{fig:spinstructure:symm_afm} are Bragg peaks that scale with the size of the system (see Figure~\ref{fig:spinscaling}) indicating long-range order, while those in~\subref{fig:spinstructure:rand_fm} and \subref{fig:spinstructure:rand_afm} are not.
$T = 0.1t$, $J/t = 3$ and the lattice contains $24 \times 24$ unit cells for each structure factor ($1152$ sites).
\label{fig:spinstructure}
}
\end{center}
\end{figure}

\begin{figure}
\begin{center}
\includegraphics[width=0.85\textwidth]{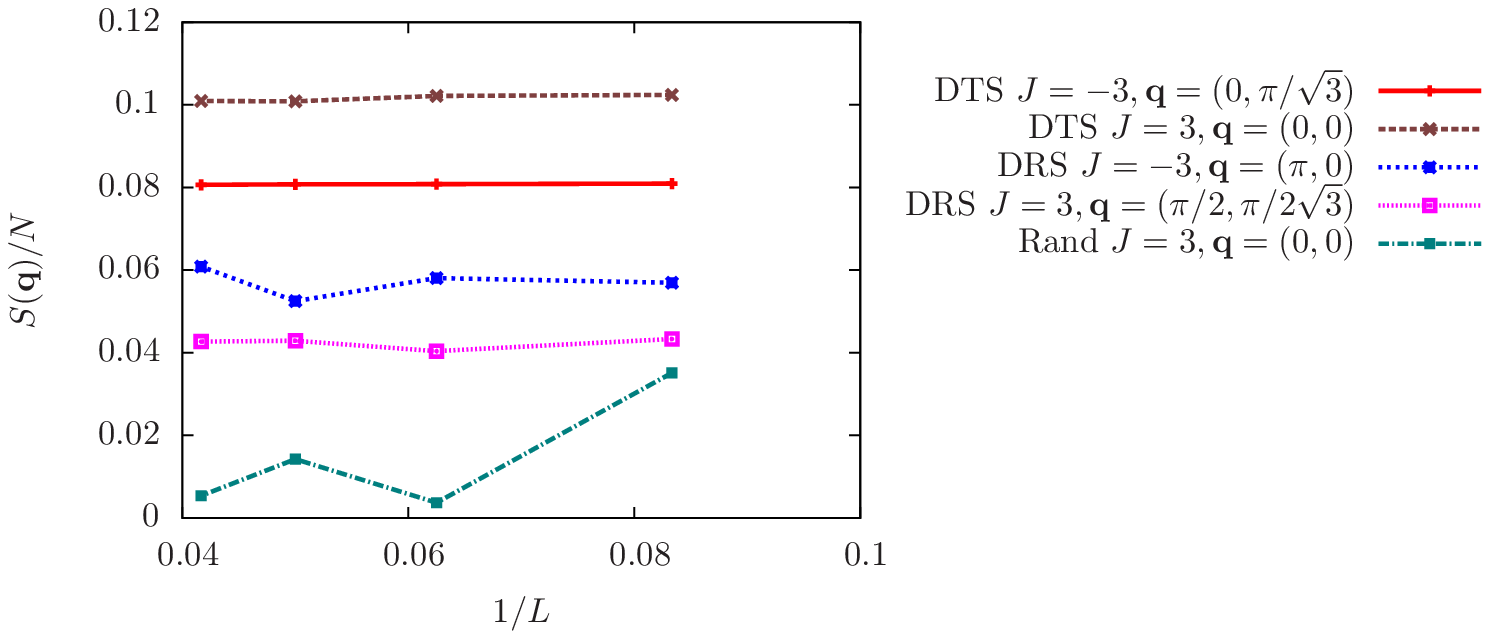}
\caption{Scaling of the peaks of the density-density structure factor for the four ordered cases and one disordered case (``random'' gauge with $J=3$).
Bragg peaks are signified by the function approaching a constant in the limit of $1/L \rightarrow 0$, where $L=\sqrt{N/2}$ is the length of the system.
Fluctuations in the scaling of the DRS gauges reflects the fact that the system has multiple ground states with distinct ordering wave vectors that must be searched, while fluctuations in the random gauge are a result of the chance presence of small clusters that support short-range correlations.
In the thermodynamic limit such clusters may result in local ordering, but do not give rise to long-range order.
\label{fig:spinscaling}
}
\end{center}
\end{figure}

Figure ~\ref{fig:spinstructure}  illustrates the density-density structure factor, calculated as
\begin{equation}
S({\bf q}) = \frac{1}{N} \sum_{ij} e^{-i {\bf q} \cdot {\bf r}_{ij}} \left< (n_i-1/2) (n_j-1/2) \right>,
\end{equation}
where ${\bf r}_{ij}$ is the vector connecting the unit cell containing site $i$ to the unit cell containing site $j$ on a honeycomb lattice and $N$ is the number of unit cells, and the sum runs over all sites of a particular sublattice.
We use the vectors between unit cells and a sublattice decomposition so that all the necessary information is contained in the familiar Brillouin zone of the triangular lattice.
Figure~\ref{fig:spinstructure} shows the density-density structure factors of both signs of $J$ for the three gauges of interest
for the case where $n_i$ and $n_j$ occur on the same sublattice.
In the case of the DTS and DRS gauges, sharp peaks occur at certain wavevectors $q_i$ in the $q$-dependent structure factor.
A peak will represent long-range order if $S(q_i)/N$ survives in the thermodynamic limit.  In Figure~\ref{fig:spinscaling}, 
we examine the finite-size scaling of the various peaks, which demonstrates that in the case of the DTS and DRS gauges,
there is indeed long-range order in the particle density.

\begin{figure}
\begin{center}
\subfigure[DTS, $J=-3$]{
\includegraphics[width=0.4\textwidth]{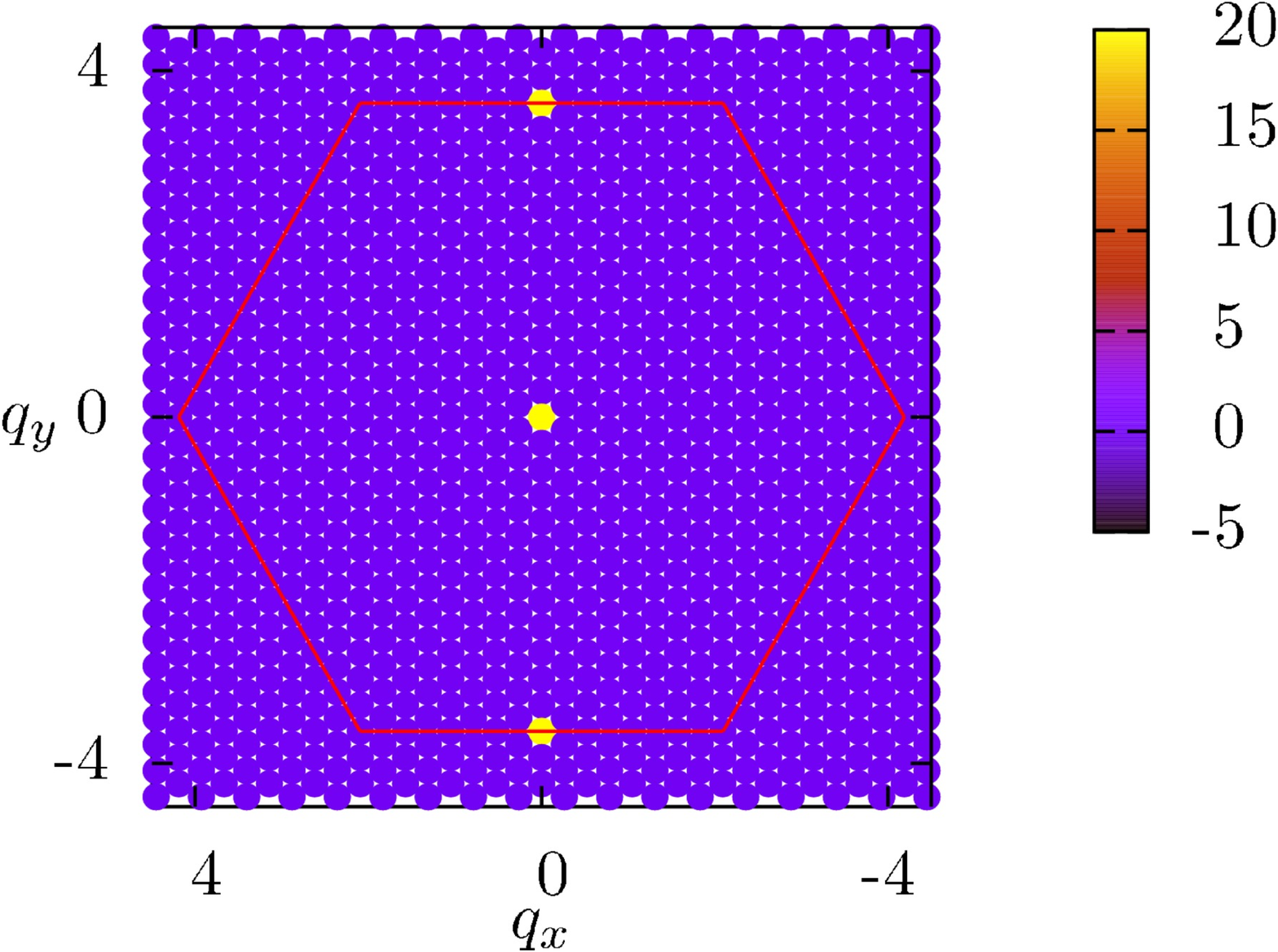}
\label{fig:bondstructure:reg_fm}
}
\subfigure[DTS, $J=3$]{
\includegraphics[width=0.4\textwidth]{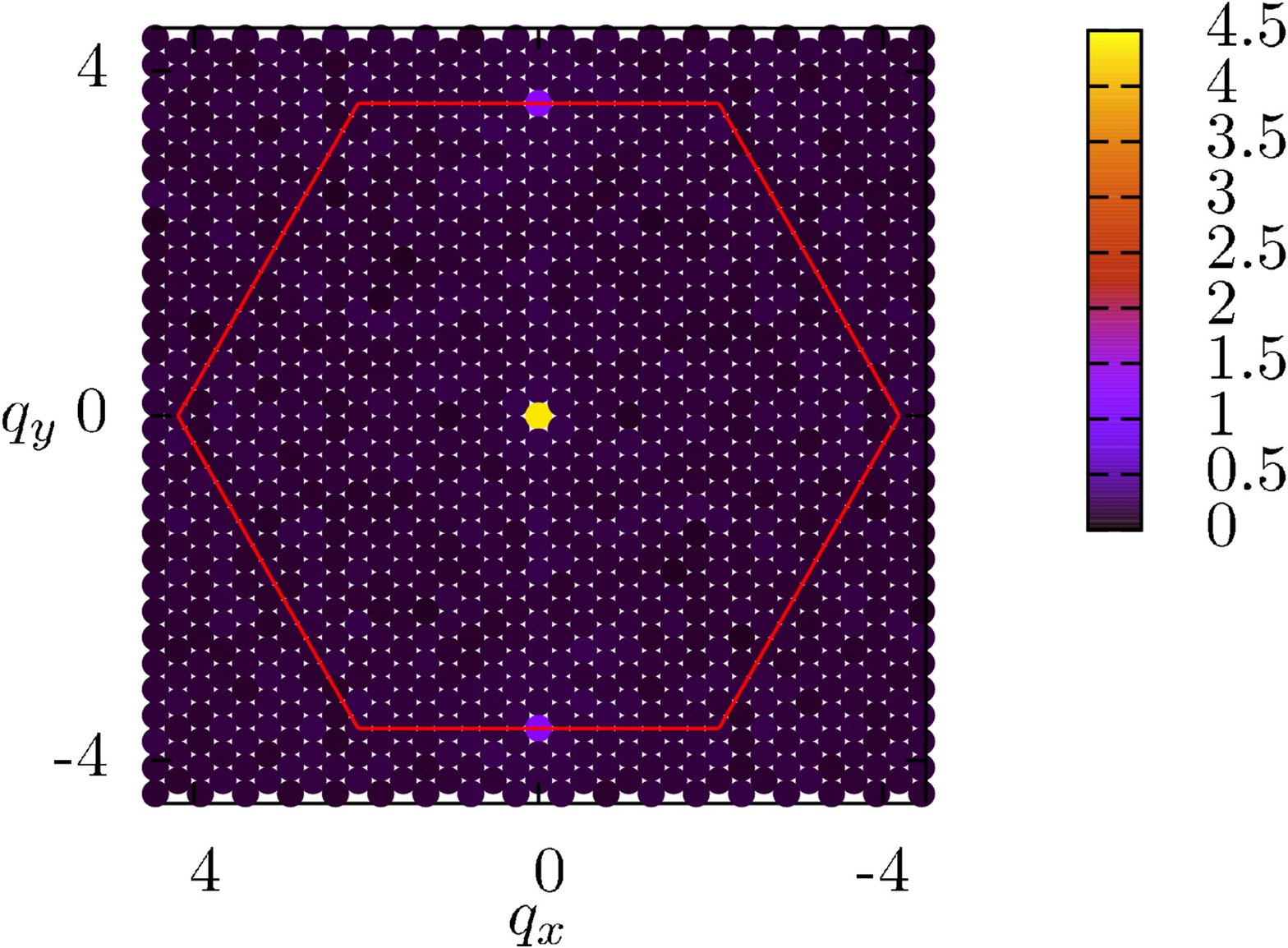}
\label{fig:bondstructure:reg_afm}
}
\subfigure[DRS, $J=-3$]{
\includegraphics[width=0.4\textwidth]{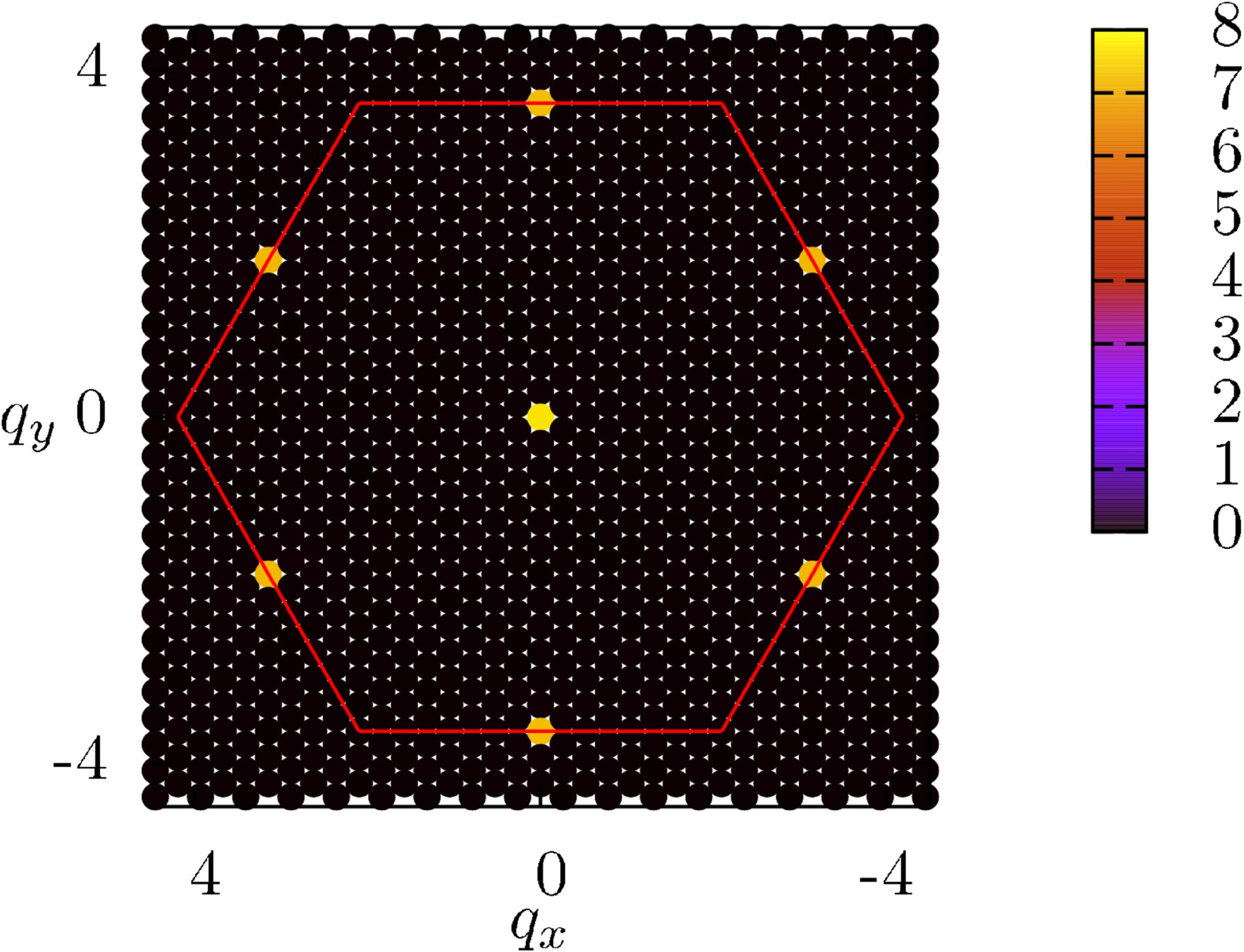}
\label{fig:bondstructure:symm_fm}
}
\subfigure[DRS, $J=3$]{
\includegraphics[width=0.4\textwidth]{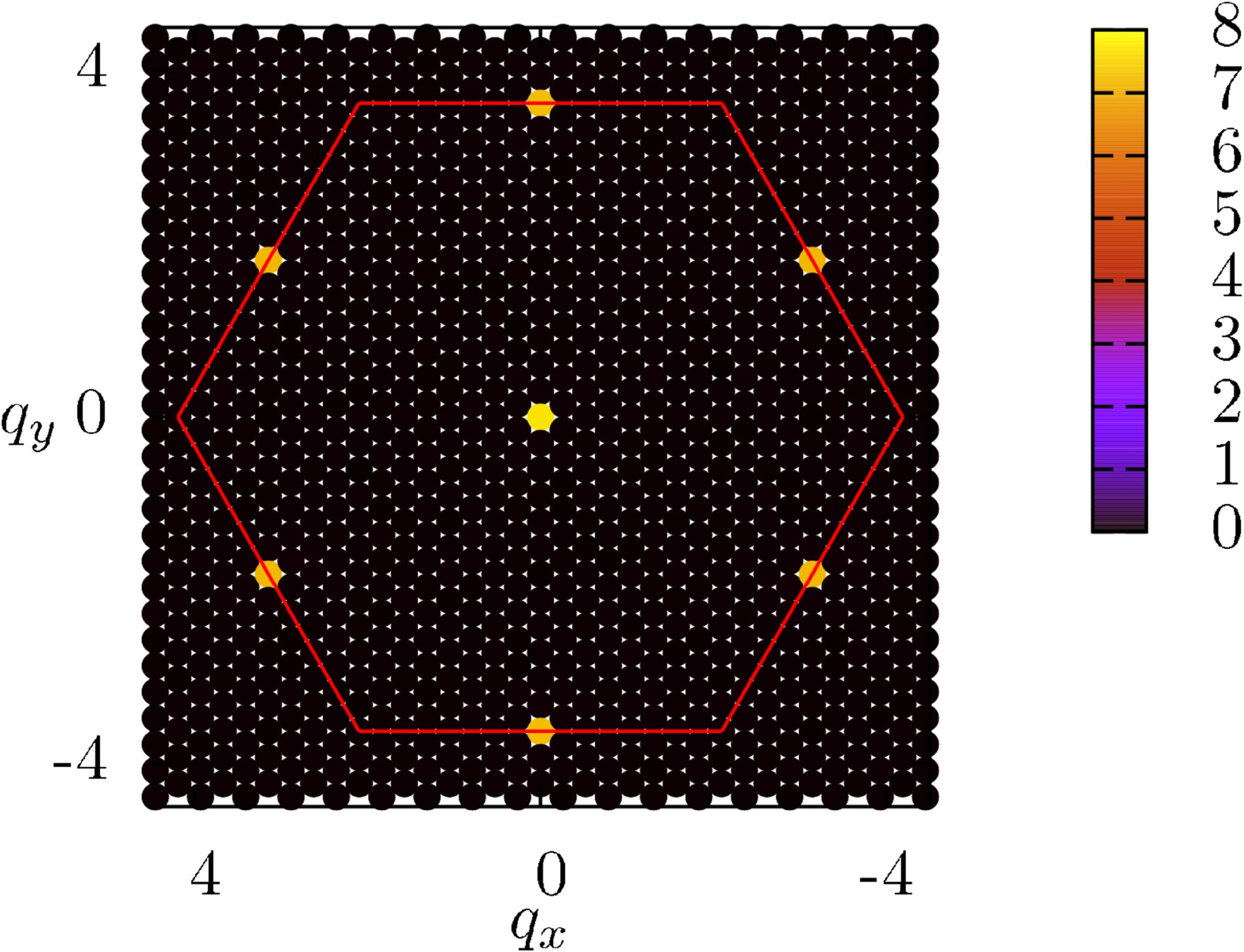}
\label{fig:bondstructure:symm_afm}
}
\subfigure[Random, $J=-3$]{
\includegraphics[width=0.4\textwidth]{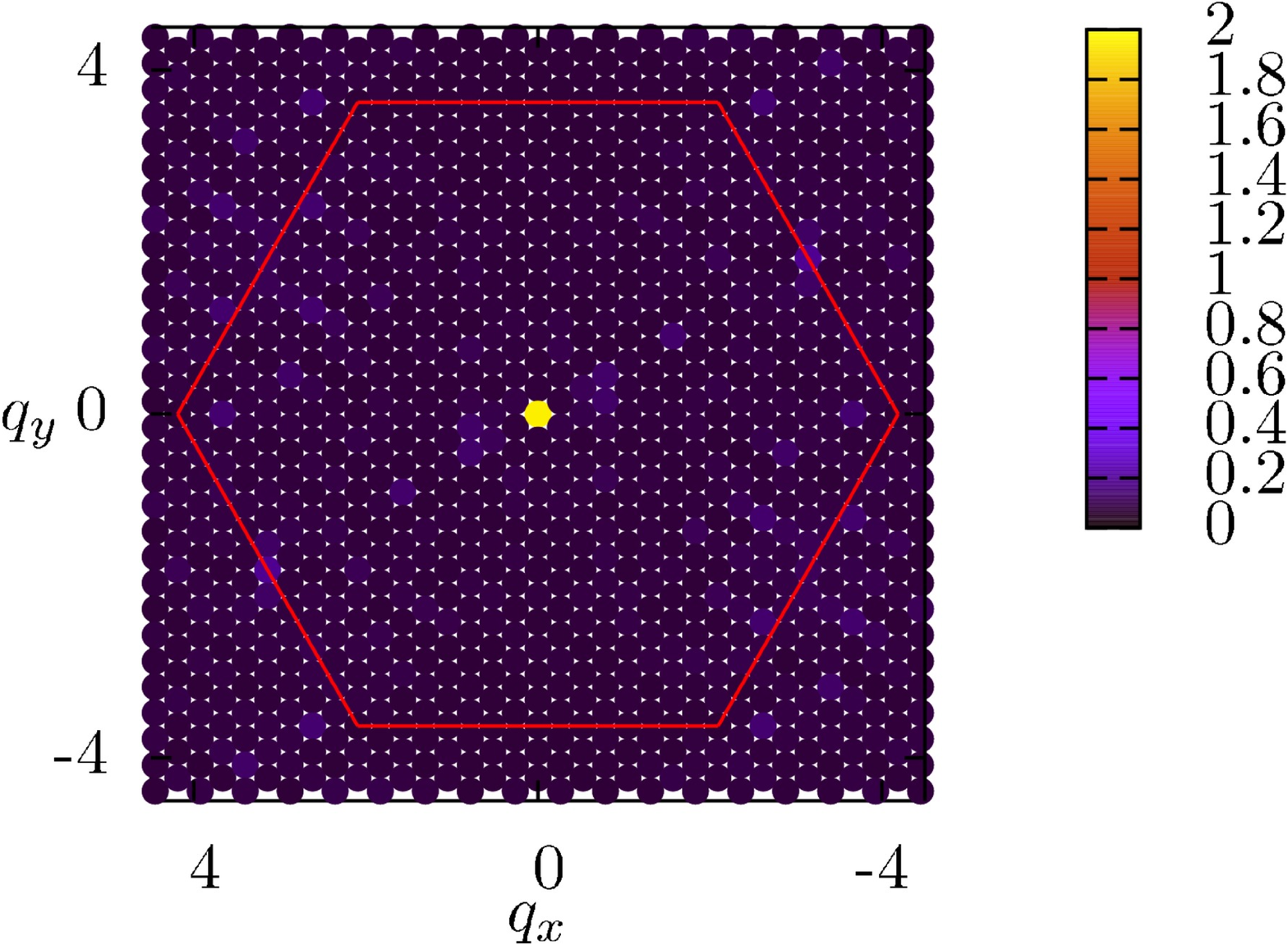}
\label{fig:bondstructure:rand_fm}
}
\subfigure[Random, $J=3$]{
\includegraphics[width=0.4\textwidth]{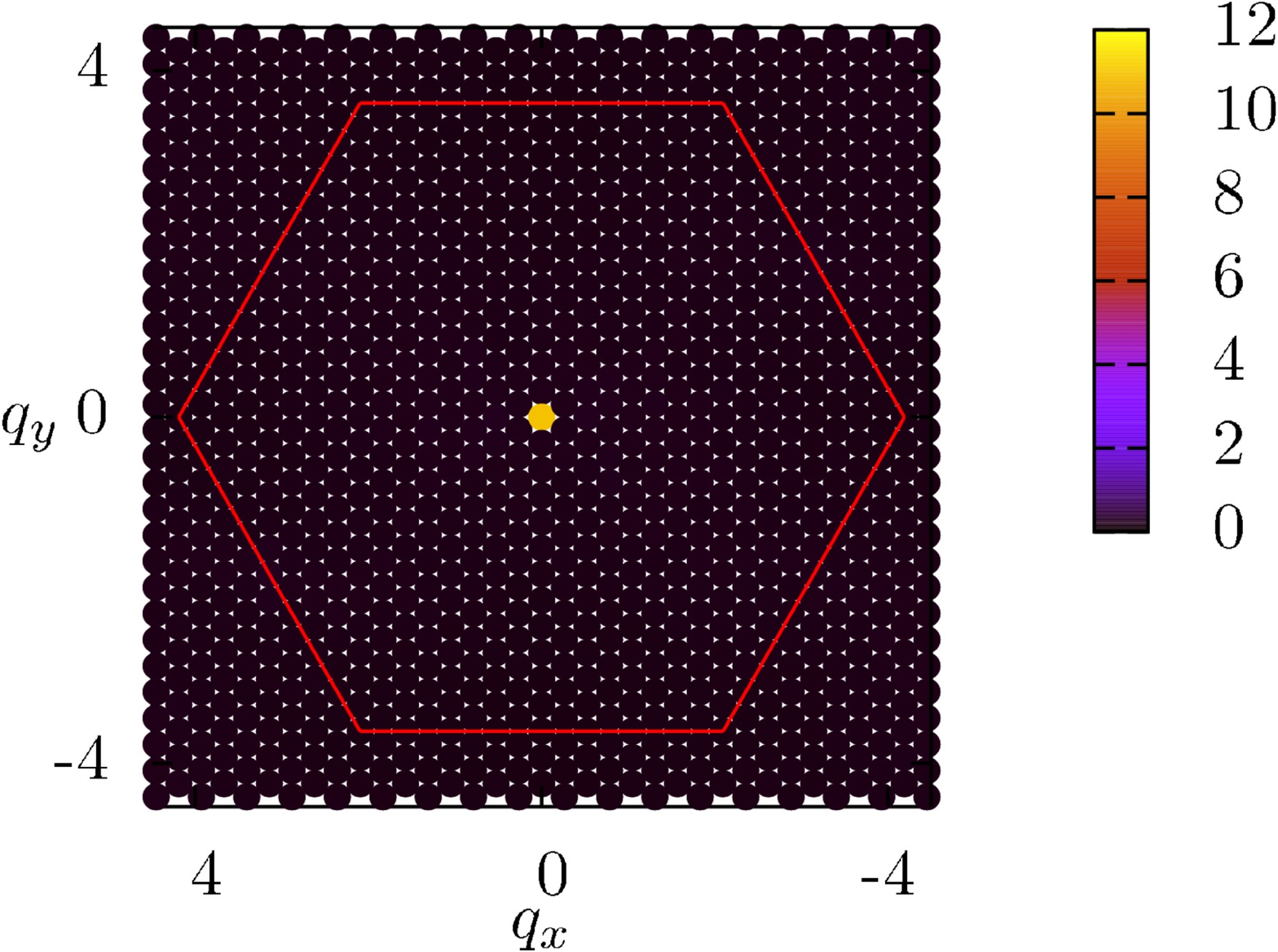}
\label{fig:bondstructure:rand_afm}
}
\caption{The structure factor of the bond-bond correlation function for the six cases studied in this paper.
The peaks in~\subref{fig:bondstructure:reg_fm},\subref{fig:bondstructure:reg_afm},\subref{fig:bondstructure:symm_fm},\subref{fig:bondstructure:symm_afm} are Bragg peaks that scale with the size of the system indicating long-range order, while those in~\subref{fig:bondstructure:rand_fm} and \subref{fig:bondstructure:rand_afm} are not.
\label{fig:bondstructure}
}
\end{center}
\end{figure}

We also measure the structure factor of the bond-bond correlation function to search for VBS order.  This structure factor is defined as
\begin{eqnarray}
S({\bf q}) &= \frac{1}{N} \sum_{ab} e^{-i {\bf q} \cdot {\bf r}_{ab}} \left< B_{a} B_{b} \right>,\\
B_{a} &= b^\dagger_i b_j + b_i b^\dagger_j, \label{hopping_def}
\end{eqnarray}
where $a$ labels the bond connecting sites $i$ and $j$ and $r_{ab}$ is the vector connecting the unit cells containing bonds $a$ and $b$.
A VBS phase will have Bragg peaks in this structure factor, which survive the thermodynamic limit as discussed above.
Figure~\ref{fig:bondstructure} shows the structure factor of the bonds for all cases studied in this paper.
Crystalline solid-like structures in the sites and the bonds are visible when the system finds in a particular symmetry broken state at low temperatures; these are the respective VBS phases.

\begin{figure}
\begin{center}
\subfigure[DTS, $J=-3$]{
\includegraphics[width=0.45\textwidth]{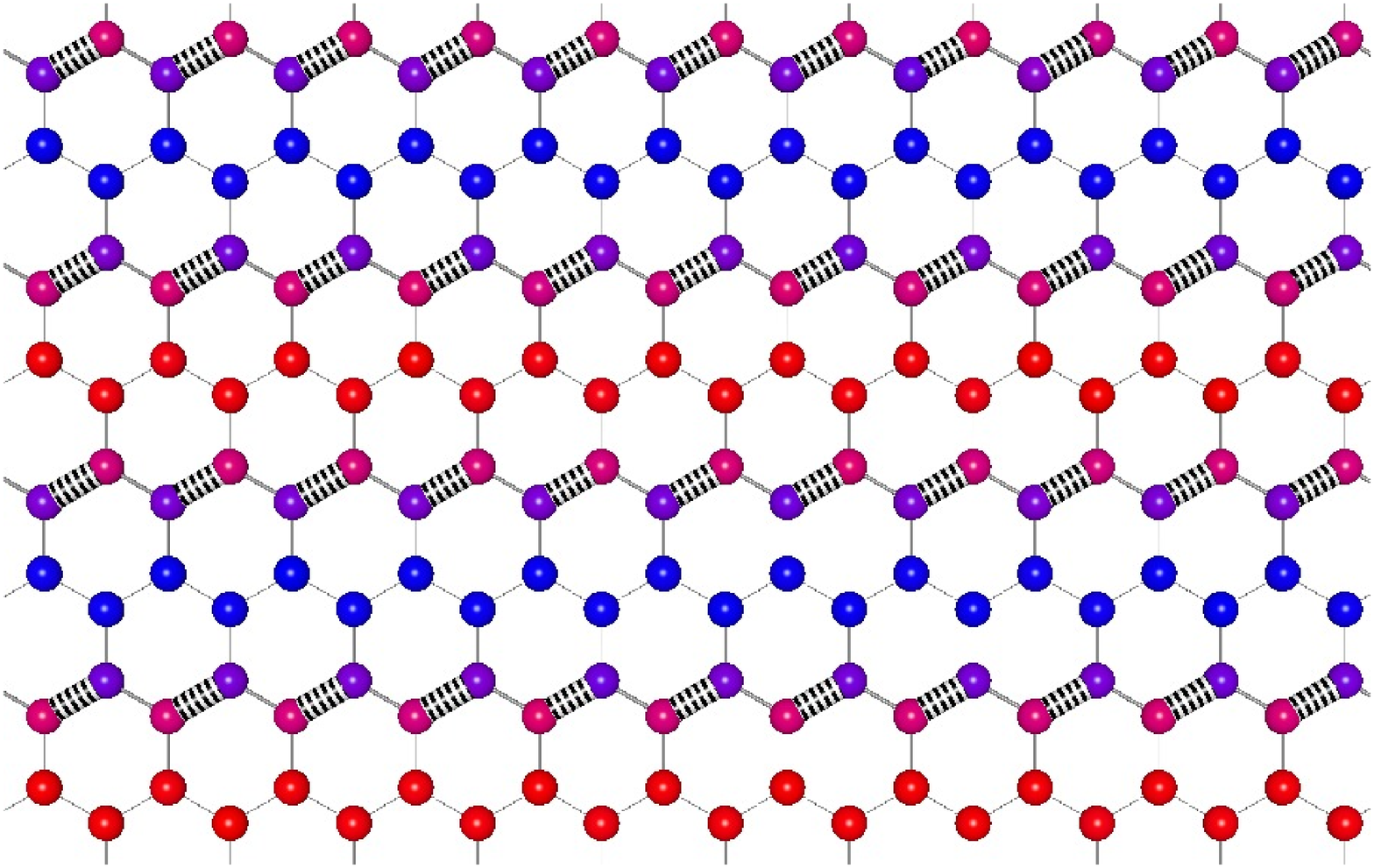}
\label{fig:rs:reg_fm}
}
\subfigure[DTS, $J=3$]{
\includegraphics[width=0.45\textwidth]{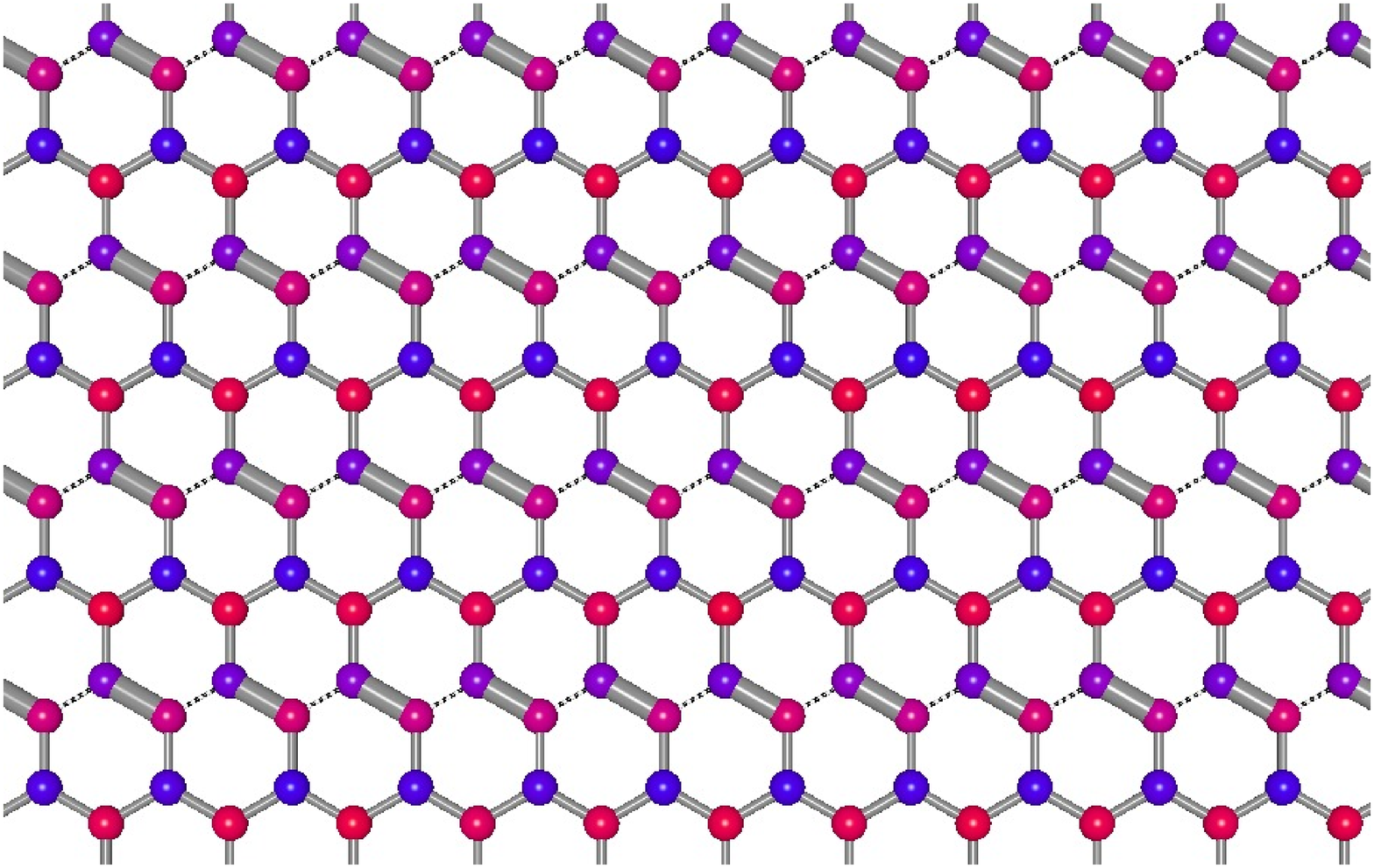}
\label{fig:rs:reg_afm}
}
\subfigure[DRS, $J=-3$]{
\includegraphics[width=0.45\textwidth]{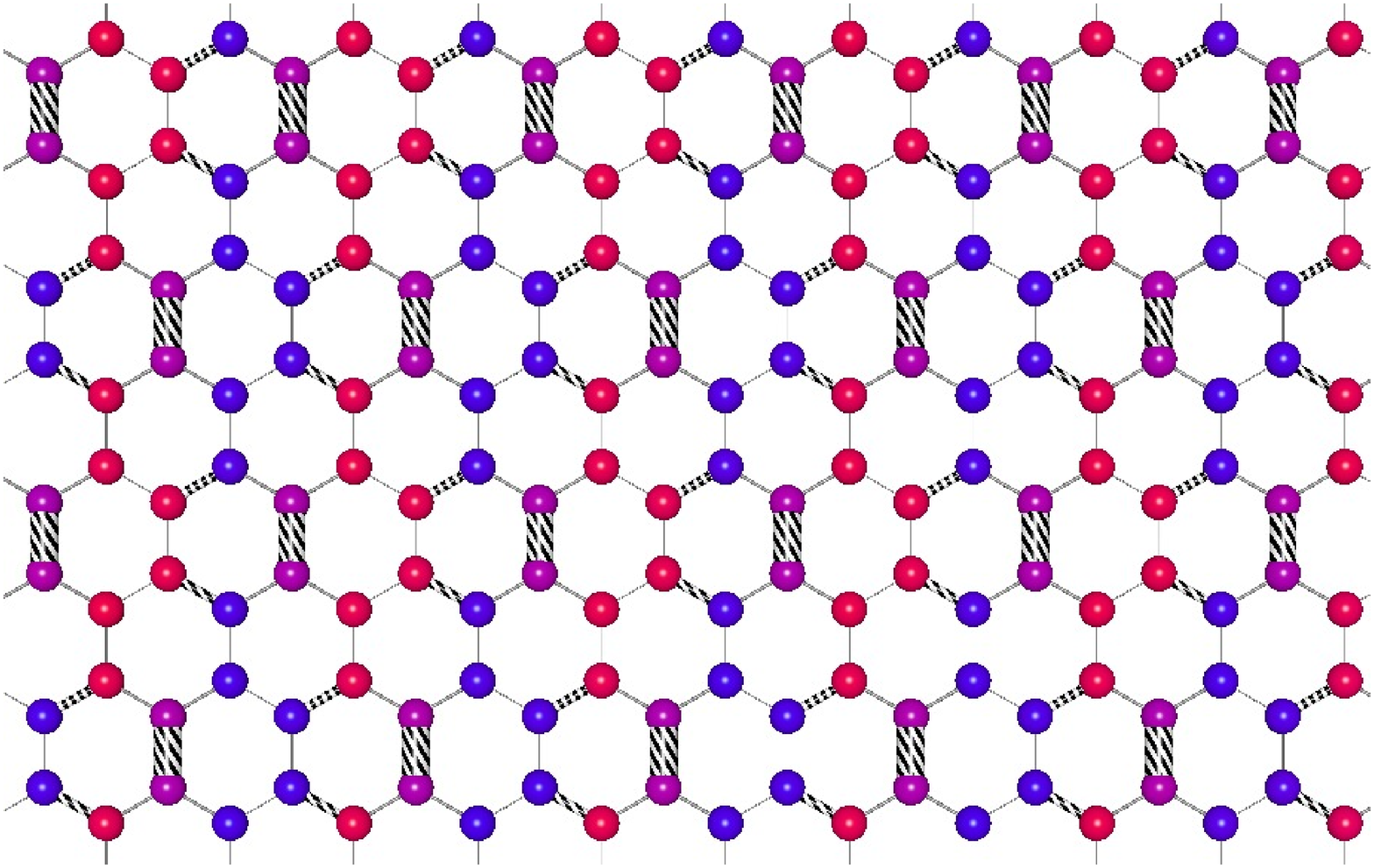}
\label{fig:rs:symm_fm}
}
\subfigure[DRS, $J=3$]{
\includegraphics[width=0.45\textwidth]{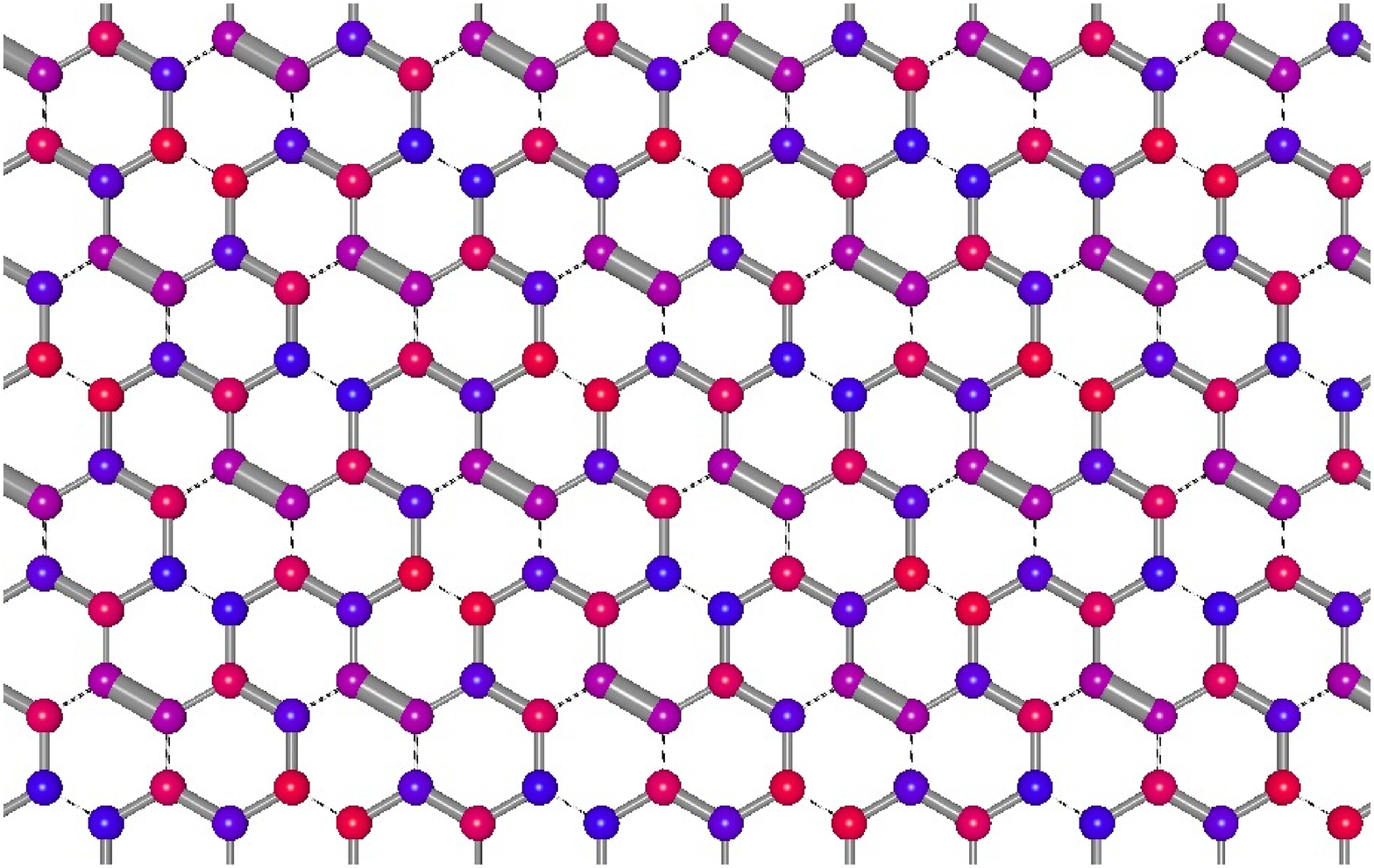}
\label{fig:rs:symm_afm}
}
\subfigure[Random, $J=-3$]{
\includegraphics[width=0.45\textwidth]{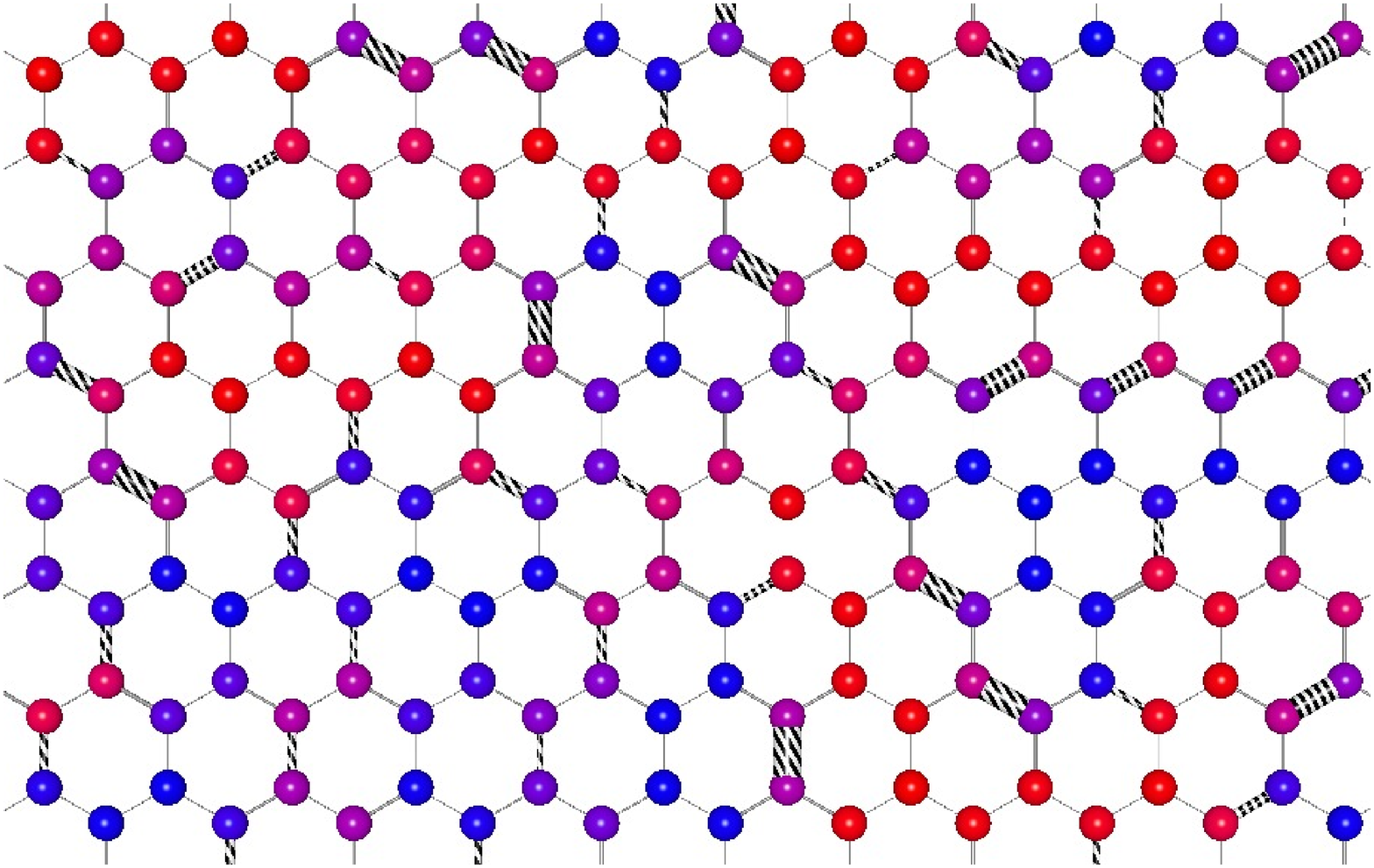}
\label{fig:rs:rand_fm}
}
\subfigure[Random, $J=3$]{
\includegraphics[width=0.45\textwidth]{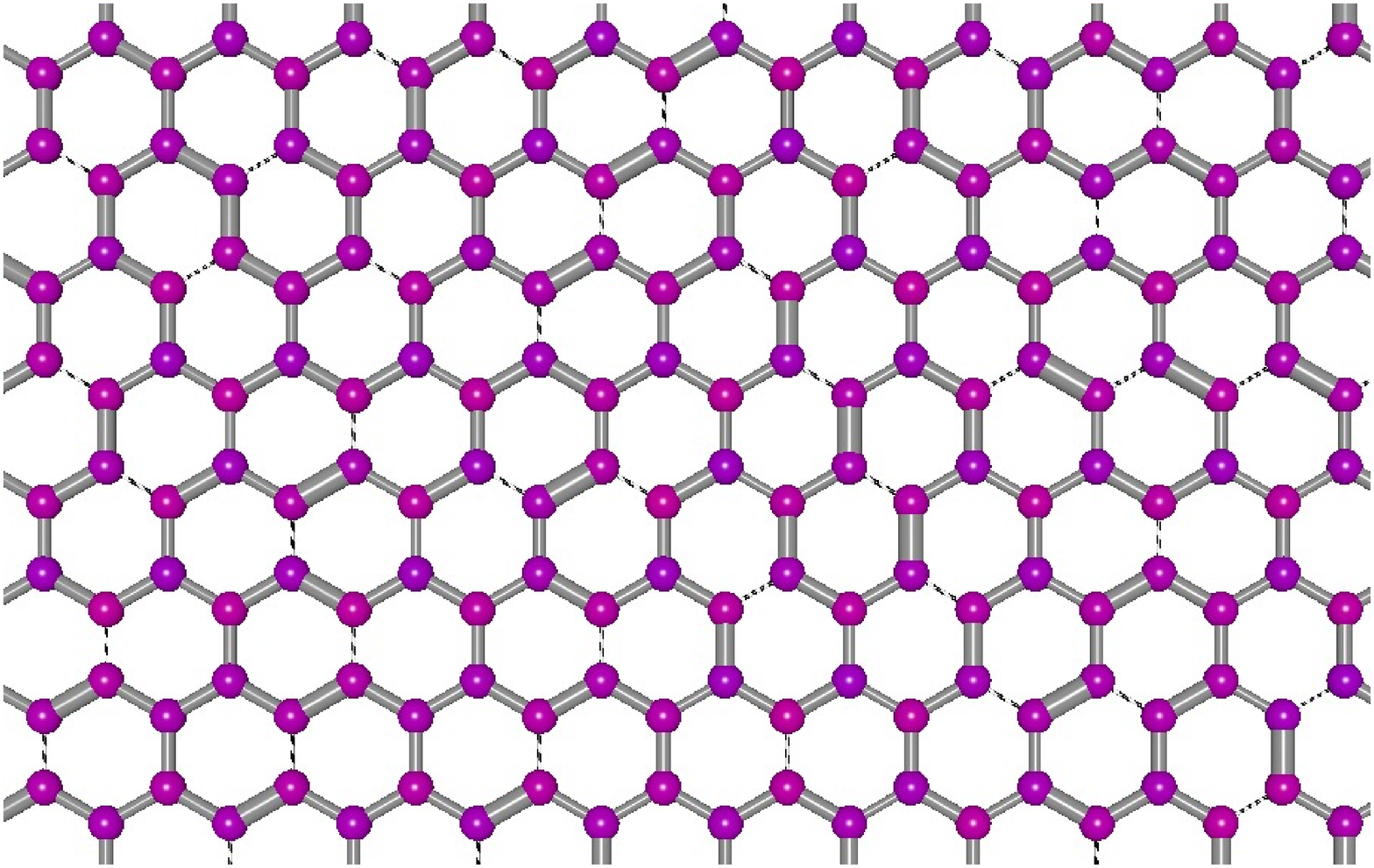}
\label{fig:rs:rand_afm}
}
\subfigure[Color-density mapping]{
\includegraphics[width=0.45\textwidth]{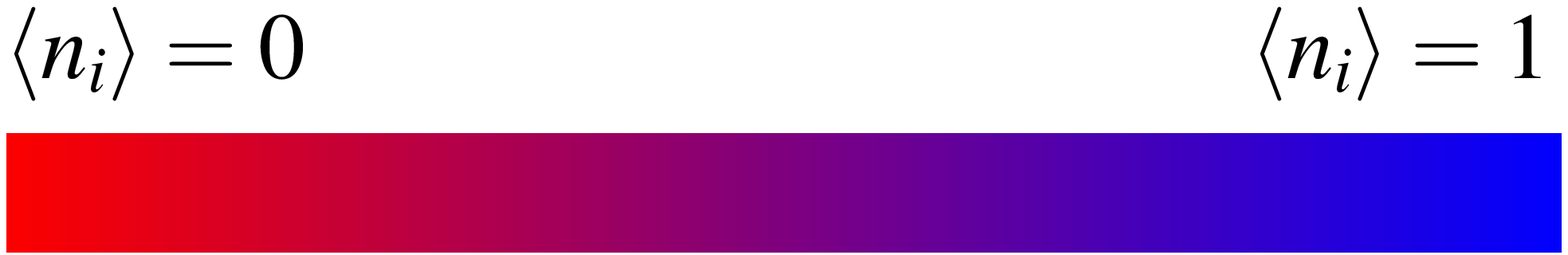}
\label{fig:rs:bar_fig}
}
\caption{The real-space image of each gauge.
Spins vary from red ($\left< n_i \right> = 0$) to blue ($\left< n_i \right> = 1$) through purple.
Bond thickness is proportional to the hopping ($\left< B_{a} \right>$), normalized to the maximum value the hopping takes during a simulation.
Striped bonds represent those where $A_{ij} = 1$.
Figures \ref{fig:rs:reg_fm}-\ref{fig:rs:symm_afm} show VBS order, while Figures \ref{fig:rs:rand_fm} and \ref{fig:rs:rand_afm} are disordered phases.
\label{fig:rs}
}
\end{center}
\end{figure}

Next, we generate real-space images of the lattice in order to help us visualize the various ground state particle and bond patterns.  In order to visualize quantum order in the 2+1 dimensional QMC simulation cell, we essentially ``project'' the SSE basis configurations and operator list back into the 2D plane - effectively averaging the simulation cell over all Monte Carlo time.  One then can imagine plotting averaged basis variables by associating a color with the expectation value of each site:  red for $\left< n_i \right> =0$, blue for $\left< n_i \right> = 1$, and mixed (shades of purple) for $0 < \left< n_i \right> < 1$ (Figure~\ref{fig:rs}).
Similarly, we average the expectation value of the kinetic energy operator $B_{a}$, defined in Equation~\ref{hopping_def}, over the imaginary time expansion, and associate a thickness of the bond
proportional to $\left< B_{a} \right>$.

The corresponding images of each gauge are presented in Figure~\ref{fig:rs}.
Figures~\ref{fig:rs:reg_fm},\ref{fig:rs:reg_afm},\ref{fig:rs:symm_fm} and~\ref{fig:rs:symm_afm} show clear long-range order in the sites and the bonds, matching up with the structure factors presented in Figure~\ref{fig:spinstructure} and Figure~\ref{fig:bondstructure}.
The lack of order in the structure factors for the random gauges is now clarified by the real-space images.
In the case where $J=-3$ the system appears to be made up of random domains of filled and empty sites, consistent with a lack of long range order.
The case where $J=3$ is the other extreme in which the system is uniformly featureless in the density, and disordered in the bonds, both again consistent with the absence of long range order.

\begin{figure}
\begin{center}
\includegraphics[width=0.75\textwidth]{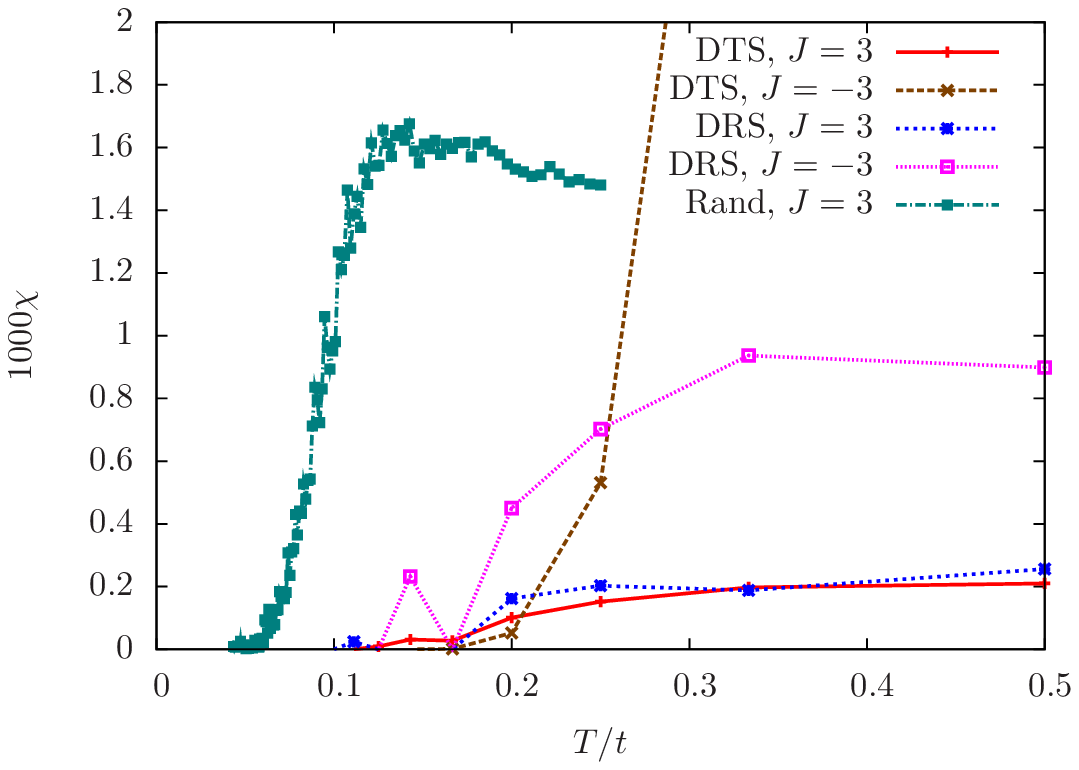}
\caption{The low temperature susceptibility of the MI in all gauges (except random $J=-3$) with $\left| J \right| = 3$.
For all cases the susceptibility approaches zero at a finite temperature, with the smallest gap in the case of the random gauge where $J=3$.
\label{fig:susc}
}
\end{center}
\end{figure}

In all four cases the average filling is half (corresponding to zero magnetization in the spin language).  The low temperature boson compressibility (uniform spin susceptibility) gives information about the density fluctuations in the groundstate: if the susceptibility goes to zero, the phase is a gapped (Mott) insulating state.
The temperature at which it turns off then gives us information about the gap to particle excitations.
In Figure~\ref{fig:susc}, susceptibility data is shown for the DTS and DRS gauges for both signs of $J$, and detailed data collected when $J=3$ for the random gauge.
For the $J=3$ random gauge, it is clear that at low temperature, the susceptibility is zero.
At high temperatures ($T \approx 0.1$), the value of the susceptibility is higher than other gauge choices.  This is likely in part because there is no long-range ordered pattern that needs to be broken to insert a particle. 
Indeed, the lack of ordering in the structure factors and real-space picture, combined with the fact that at low-temperature the phase is incompressible, suggests that the state is a featureless Mott insulating state - a candidate quantum spin liquid.

%% file: conclusions.tex
\pagebreak
\section{Quantum Order by Disorder}

\begin{figure}
\begin{center}
\includegraphics[width=0.45\textwidth]{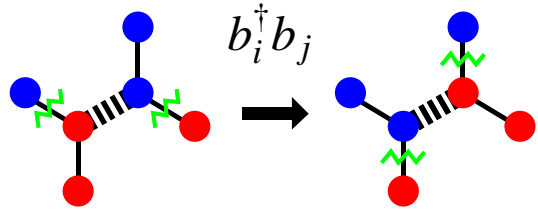}
\caption{The left state is taken to the right state by the action of the $b^\dagger_i b_j$ element of the Hamiltonian.
If the left state satisfies the classical ($J$) part of the Hamiltonian, so will the right state.
In this way, the superposition of these two (local) states reduces the energy of the quantum ($t$) Hamiltonian without increasing energy in the classical part of the Hamiltonian.
Blue (red) circles represent particles (holes), while the green zigzags represent frustrated bonds.
\label{fig:perturb}
}
\end{center}
\end{figure}

The results from QMC show a superfluid-MI phase transition in each of the cases studied, where the Mott insulator is a symmetry-broken crystalline state whenever the gauge choice is repeating, and a featureless state when the gauge is random.  
Keeping in mind that the $t$ term of the Hamiltonian is off-diagonal in the density-basis choice, we can understand the crystalline phases in the limit of large $J/t$.
Namely, treating $t$ as a perturbation on top of the classical Ising model degenerate groundstate, one can ask which local $t$ operations
can lower the kinetic energy (maximize hopping) without costing energy proportional to $J$.  In all cases we can see that the crystalline (VBS) states discovered by the QMC satisfy these two criteria.  The simplest case 
comes from the DTS gauge where $J=-3$.
If we look where the expectation of the hopping operator is largest, it occurs on bonds where energy is minimized when a particle is adjacent to a hole.
The two hopping sites are also connected to two frustrated bonds before and after exchanging their positions, as shown in Figure~\ref{fig:perturb}.
The real-space figure of the ground state shows a mixing of the two states in Figure~\ref{fig:perturb}, suggesting that such a superposition is also energetically favorable in the simulation.

In the case where $J=-3$ in the manifold of states that satisfy the classical part of the Hamiltonian, the hopping operator will only act on bonds that prefer holes next to particles.
This is because swapping such sites is the only way to remain in the set of states that satisfies the classical part of the Hamiltonian.
The only scenario in which the hopping operator can act on an attractive bond is if it is frustrated.
Looking at Figure~\ref{fig:perturb} again, all the bonds connected to the two swapped sites, except the bond between them, switch from frustrated to unfrustrated and vice versa when the sites are swapped.
\begin{figure}
\begin{center}
\includegraphics[width=0.5\textwidth]{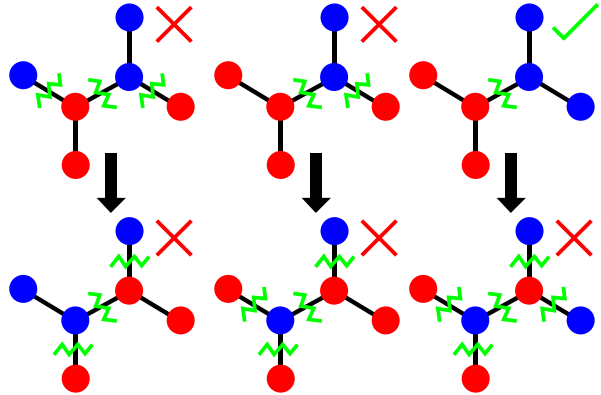}
\caption{When the central bond is a frustrated attractive bond, all action of the hopping operator has a state that costs and energy of order $J$ (denoted by an ``X'') as either the state before and/or after hopping.
Blue (red) circles represent particles (holes), while the green zigzags represent frustrated bonds.
\label{fig:no_fm}
}
\end{center}
\end{figure}
Figure~\ref{fig:no_fm} shows how there is no configuration possible that would allow hopping on an attractive bond to be a part of the ground state.

\begin{figure}
\begin{center}
\includegraphics[width=0.55\textwidth]{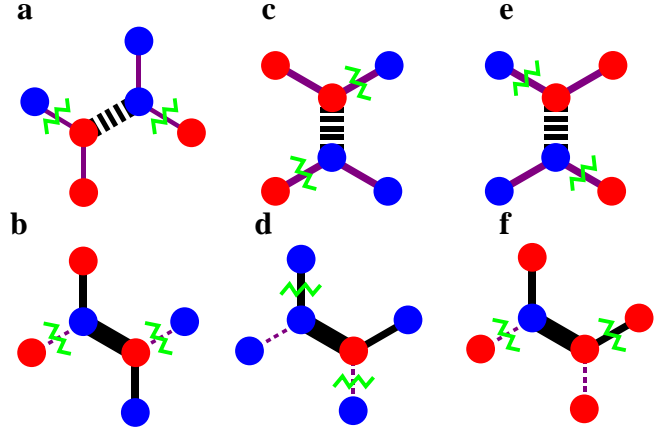}
\caption{A comparison of the local structure around highly fluctuating bonds from all gauges.
In the top row of images $J=-3$, so striped bonds represent repulsive bonds.
In the bottom row, $J=3$, so striped bonds represent attractive bonds.
Blue (red) circles represent particles (holes), while the green zigzags represent frustrated bonds.
Due to both gauges being presented, attractive bonds have been noted in purple.
The top layer shows the ({\bf a}) DTS gauge and two ({\bf c},{\bf e}) local configurations of the DRS gauge, all with $J=-3$.
The bottom layer shows the ({\bf b}) DTS gauge and again two ({\bf d},{\bf f}) local configurations of the DRS gauge, all with $J=3$.
\label{fig:perturb_compare}
}
\end{center}
\end{figure}

By eliminating the possibility that the attractive bonds fluctuate to lower the system's energy we find that only the repulsive bonds will, and the numerical simulations confirm this result.
The ground state can now be described as the superposition of states that allows each repulsive bond to fluctuate as much as possible.
If we extend this simple idea to all the gauges and signs of $J$, we can check to see if the highly fluctuating bonds in each case correspond to local configurations that satisfy configurations similar to Figure~\ref{fig:perturb}.
Figure~\ref{fig:perturb_compare} shows the local structure around each of the highly fluctuating bonds from the DTS and DRS gauges, and in each case the local configuration satisfies what we expect from the perturbation theory approach.

\begin{figure}
\begin{center}
\includegraphics[width=0.5\textwidth]{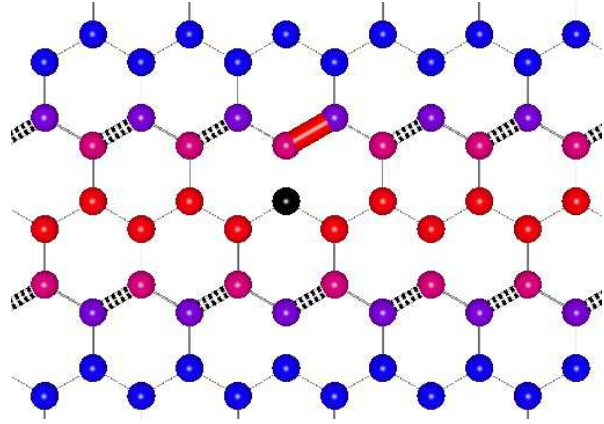}
\caption{A real-space picture of the lattice generated using the bond-bond and density-density correlation function, with the red bond and black site as the reference bond.
Notice the subtle oscillation in bond thickness for those bonds on the same layer as the reference.
\label{fig:bond_pic}
}
\end{center}
\end{figure}

\begin{figure}
\begin{center}
\includegraphics[width=0.6\textwidth]{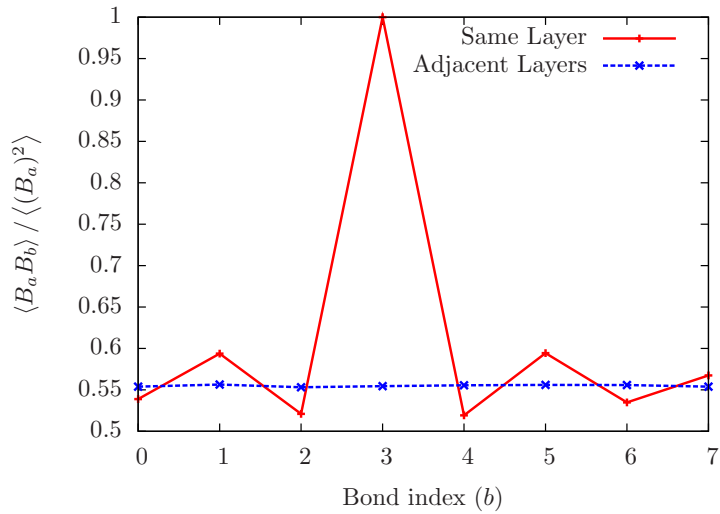}
\caption{Numerical version of the data presented in Figure~\ref{fig:bond_pic}, with $a = 3$.
Here the oscillation in the correlation function is very clear, with anti-correlation of nearest neighbors.
\label{fig:bond_numeric}
}
\end{center}
\end{figure}

To further cement our understanding of the fluctuations, we can use the information extracted from QMC measurement of the bond-bond correlation function, $\left< B_{a} B_{b} \right>$, to see if it matches these expectations.
We'll use the simplest case to build our intuition, the DTS gauge with $J=-3$.
In this case when looking at a layer of highly fluctuating bonds in their dominant configuration, any of them may hop to reduce energy.
Once one of them hops however, neighbors on the same layer that were able to hop before are no longer able to.
When we compare this with numerical results, we find this nearest neighbor behavior is reproduced.
Figure~\ref{fig:bond_pic} shows the real-space image of the ground state generated using the bond-bond correlation function.
Notice how the thickness of the bonds next to the reference bond (red), corresponding to $\left< B_{a} B_{b} \right>$, is thinner for those adjacent compared to those two unit cells away.
The graph in Figure~\ref{fig:bond_numeric} illustrates the thickness as a function of position along the chain, and clearly shows that adjacent bonds are anti-correlated when compared to well separated bonds.

\section{Discussion}

From our QMC data we have demonstrated that a superfluid-Mott insulator phase transition can exist in the nearest-neighbor Bose Hubbard model on the honeycomb lattice where the interactions are constrained to be fully-frustrating around each hexagon, subject to a gauge choice.  We are able to classify the Mott insulating phases in each case where the gauge is a repeating pattern as a symmetry-broken crystal with coexisting charge-density wave and valence-bond order (i.e.~a CDW-VBS~\cite{jinwuye_cdw}).  Bragg peaks in the density-density and bond-bond correlations functions confirm the long-range order that we have illustrated using a real-space averaging of the QMC simulation cell.  One could describe the Mott insulating phases as occurring by an order-by-disorder mechanism, where the degeneracy of the classical manifold of states for $t=0$ is lifted by the quantum ``perturbation'' or hopping.  We have intuitively described this order-by-disorder mechanism by a simple perturbation theory picture to first order in the hopping, subject to the constraint that each hopping process costs no energy in the (classical) interaction.

When the gauge choice of the interaction frustration is {\it not} a repeating pattern, we find that the order-by-disorder mechanism is destroyed.  In this case, QMC finds a superfluid-Mott insulating transition, where the Mott insulator is a disordered phase, featureless on long length-scales.  Both the $q$-space structure factors and the real-space simulation cell pictures confirm the groundstate configurations have no long range order. For $J=-3$ (predominantly attractive interactions),
repeated simulations show that the configuration of the ground state consists of domains of filled or empty sites.
In this way, the random gauge with $J=-3$ can be considered a disordered solid in the density and valence bond sectors.
Since most of the hopping dynamics take place on the interface between domains, excitations of this gauge should be similar to those in the DTS gauge when $J=-3$.

The random gauge with $J=3$ is the most interesting and unusual case, lacking order in the density and displaying a ground state where every site is exactly half filling on average.
The density-density correlation function quickly decays to zero (within error bars) in a few lattice spacings.
Like all the other gauges, the random gauge also has a gap to single particle excitations, shown by the susceptibility reducing to zero at low temperatures.  Despite the gauge choice breaking all lattice symmetries, this phase is {\it not} described by a Bose-glass, since it is gapped with a vanishing local density Edwards-Anderson (EA) order parameter.
One should also note that although the MI has no long-range order, it does have a somewhat broad distribution of bond strength (as evident in the differences in bond thickness in Figure \ref{fig:rs}(f)), which is induced locally due to the random pattern of frustration.  
This local disorder can be thought of as occurring ``on top'' of the RVB resonances.
One may also imagine a different phase with no EA order parameter but disordered valence bonds, where spins are coupled in randomly frozen {\it pairs} (singlets) --  a so-called Valence Bond Glass \cite{RajivVBglass}.  In contrast, spins in our phase are not coupled into singlets but free to resonate around the lattice, suggesting its description as a ``glassy'' RVB phase. 

Thus, one can consider this MI is a good candidate for a quantum spin liquid groundstate -- a state which is often defined as a $T=0$ disordered phase that has an emergent gauge symmetry, associated with fractional excitations \cite{balents2010spin}.
It is therefore interesting to consider briefly the excitations out of the groundstate.
In the limit of $t = 0$, the only excitations in our model correspond to inserting a single particle and raising the energy of the system by $J$.
In the limit of small $t/J$, different superpositions of classically allowed states will have different expectations of energy, but finding eigenstates other than the ground state is exceedingly difficult in SSE QMC.
Since the total particle number commutes with the Hamiltonian, we can use it as a quantum number to label different states.
If we can change the total filling of our ground state without violating the classical constraint, the new state we have can only be different in energy by some polynomial of $t$, or it may have a zero energy difference.
The earlier discussion of hopping taken with attractive bonds leads us to one method of constructing such a state.

Let us imagine a situation similar to Figure~\ref{fig:perturb}, except with the central bond taken to be a satisfied attractive bond.
In this scenario if we insert or remove two particles the number of frustrated bonds does not change, but we are taken to new eigenstate of the quantum Hamiltonian.
Using QMC simulations with an applied chemical potential we were able to explore the excited state with two extra particles, compared to a half filled system.
The evolution of the simulation suggests a large barrier between the half filled state and the state with two extra particles, as the state with one extra particle must have an energy penalty of order $J$, consistent with our earlier discussion.
Repeated simulations from random initial configurations show us that states away from half filling must have some energy penalty, as the simulation always finds states with half filling at low temperatures as the lowest energy state.
This implies that there is an energy difference between the ground state and that with two extra particles, and the difference between these states is not of order $J$, but some order of $t$.  It is therefore interesting that the lowest-energy excitations in this groundstate do not appear to be the typical vison and spinon expected of a Z2 spin liquid \cite{isakov_spinliquid}.  Further characterization of the excitations in the model will be needed to make a possibles connection to a theoretical description of a quantum spin liquid phase.

In conclusion, we have studied a fully-frustrated Bose Hubbard model on the honeycomb lattice, where the frustration is induced by explicitly modulating the sign of the interaction around each hexagon to preclude a satisfied interaction.  This choice of (classical) gauge can induce order-by-disorder when perturbed by the quantum hopping term, in the case when the gauge choice has a regular spatial pattern.  In the case where the gauge choice is explicitly chosen to break all lattice symmetries, a disordered Mott insulating or RVB state is induced.  Further work to look at excitations in this phase may be needed in order to determine its suitability as a quantum spin liquid state, i.e.~a
phase with fractionalized excitations and emergent gauge symmetry.  Although more refinement of the model may be necessary, the identification of such candidate spin liquid states in Bose Hubbard Hamiltonians with only nearest-neighbor interactions is an important step towards the creation of such phases in cold atomic systems in optical lattices.

\section{Acknowledgements}

We are grateful for enlightening discussions with R. Singh, F. Heidrich-Meisner, L.~Balents, Y-B  Kim, R. Moessner, S. Isakov and A. Burkov.  Simulations were performed using the computing facilities of SHARCNET. This work was supported by NSERC of Canada (S.I. and R.G.M.), Vanier CGS (S.I.) and the National Science Foundation under grant NSF PHY05-51164 (R.G.M.)

\section{References}